\documentclass[]{aastex631}
\usepackage[utf8]{inputenc}

\usepackage{tikz}
\usetikzlibrary{calc,tikzmark}

\usepackage{url}
\usepackage{xspace}
\usepackage{upgreek}
\usepackage{booktabs}

\newcommand{\msun}{\ensuremath{\mathrm{M}_\odot}\xspace}
\newcommand{\lsun}{\ensuremath{\mathrm{L}_\odot}\xspace}
\newcommand{\kms}{\ensuremath{\mathrm{km~s}^{-1}}\xspace}
\newcommand{\um}{\ensuremath{\mathrm{\mu m}}\xspace}
\newcommand{\percc}{\ensuremath{\mathrm{cm}^{-3}}\xspace}
\newcommand{\persc}{\ensuremath{\mathrm{cm}^{-2}}\xspace}
\newcommand{\pers}{\ensuremath{\mathrm{s}^{-1}}\xspace}
\def\ee#1{\ensuremath{\times10^{#1}}}
\newcommand{\threemm}{3~mm\xspace}
\renewcommand{\deg}{\ensuremath{^{\circ}}\xspace}
\def\rr#1{{#1}}

\newcommand{\hcopthirt}{\ensuremath{\mathrm{H}^{13}\mathrm{CO}^+}\xspace}

\begin{document}

\title{A broad linewidth, compact, millimeter-bright molecular emission line source near the Galactic Center}
\author[0000-0001-6431-9633]{Adam Ginsburg}
\affiliation{Department of Astronomy, University of Florida, P.O. Box 112055, Gainesville, FL 32611}

\author[0000-0001-8135-6612]{John Bally} 
\affiliation{Center for Astrophysics and Space Astronomy, 
     Department of Astrophysical and Planetary Sciences \\
     University of Colorado, Boulder, CO 80389, USA}

\author[0000-0003-0410-4504]{Ashley~T.~Barnes}
\affiliation{European Southern Observatory (ESO), Karl-Schwarzschild-Stra{\ss}e 2, 85748 Garching, Germany}

\author[0000-0002-6073-9320]{Cara Battersby}
\affiliation{Department of Physics, University of Connecticut, 196A Auditorium Road, Unit 3046, Storrs, CT 06269, USA}

\author[0000-0002-0533-8575]{Nazar Budaiev}
\affiliation{Department of Astronomy, University of Florida, P.O. Box 112055, Gainesville, FL 32611}

\author[0000-0002-4013-6469]{Natalie O. Butterfield}
\affiliation{National Radio Astronomy Observatory, 520 Edgemont Road, Charlottesville, VA 22903, USA}

\author[0000-0003-1481-7911]{Paola Caselli}
\affiliation{Max-Planck-Institut f\"ur extraterrestrische Physik, Giessenbachstrasse 1, D-85748 Garching, Germany}

\author[0000-0001-8064-6394]{Laura Colzi}
\affiliation{Centro de Astrobiología (CAB), CSIC-INTA, Carretera de Ajalvir km 4, Torrejón de Ardoz, 28850 Madrid, Spain}

\author[0000-0003-0980-6871]{Katarzyna M. Dutkowska}
\affiliation{Leiden Observatory, Leiden University, P.O. Box 9513, 2300 RA Leiden, The Netherlands}

\author[0000-0002-8586-6721]{Pablo Garc\'ia}
\affiliation{Instituto de Astronomía, Universidad Católica del Norte, Av.\ Angamos 0610, Antofagasta, Chile}
\affiliation{Chinese Academy of Sciences South America Center for Astronomy, National Astronomical Observatories, CAS, Beijing 100101, China}

\author[0000-0002-1313-429X]{Savannah Gramze}
\affiliation{Department of Astronomy, University of Florida, P.O. Box 112055, Gainesville, FL 32611}

\author[0000-0001-9656-7682]{Jonathan D. Henshaw}
\affiliation{Astrophysics Research Institute, Liverpool John Moores University, IC2, Liverpool Science Park, 146 Brownlow Hill, Liverpool L3 5RF, UK}
\affiliation{Max Planck Institute for Astronomy, K\"{o}nigstuhl 17, D-69117 Heidelberg, Germany}

\author[0000-0002-8455-0805]{Yue Hu}
\affiliation{Department of Physics, University of Wisconsin-Madison, Madison, WI, 53706, USA}

\author[0000-0003-0416-4830]{Desmond Jeff}
\affiliation{Department of Astronomy, University of Florida, P.O. Box 112055, Gainesville, FL 32611}
\affiliation{National Radio Astronomy Observatory, 520 Edgemont Road, Charlottesville, VA 22903-2475, USA}

\author[0000-0003-4493-8714]{Izaskun Jim\'enez-Serra}
\affiliation{Centro de Astrobiología (CAB), CSIC-INTA, Carretera de Ajalvir km 4, Torrejón de Ardoz, 28850 Madrid, Spain}

\author[0000-0002-5094-6393]{Jens Kauffmann}
\affiliation{Haystack Observatory, Massachusetts Institute of Technology, 99 Milsstone Road, Westford, MA 01886, USA}

\author[0000-0002-0560-3172]{Ralf S.\ Klessen}
\affiliation{Universit\"at Heidelberg, Zentrum f\"ur Astronomie, Institut f\"ur Theoretische Astrophysik, Albert-Ueberle-Str. 2, 69120 Heidelberg, Germany}
\affiliation{Universit\"{a}t Heidelberg, Interdisziplin\"{a}res Zentrum f\"{u}r Wissenschaftliches Rechnen, Im Neuenheimer Feld 205, 69120 Heidelberg, Germany}

\author[0000-0003-2184-1581]{Emily M. Levesque}
\affiliation{Department of Astronomy, Box 351580, University of Washington, Seattle, WA 98195, USA}

\author[0000-0001-6353-0170]{Steven N. Longmore}
\affiliation{Astrophysics Research Institute, Liverpool John Moores University, IC2, Liverpool Science Park, 146 Brownlow Hill, Liverpool L3 5RF, UK}
\affiliation{COOL Research DAO}

\author[0000-0003-2619-9305]{Xing Lu}
\affiliation{Shanghai Astronomical Observatory, Chinese Academy of Sciences, 80 Nandan Road, Shanghai 200030, People's Republic of China}

\author[0000-0001-8782-1992]{Elisabeth A.C. Mills}
\affiliation{Department of Physics and Astronomy, University of Kansas, 1251 Wescoe Hall Drive, Lawrence, KS 66045, USA}

\author[0000-0002-6753-2066]{Mark R. Morris}
\affiliation{Department of Physics \& Astronomy, University of California, Los Angeles, Los Angeles, CA 90095-1547, USA}

\author[0000-0002-6379-7593]{Francisco Nogueras-Lara}
\affiliation{European Southern Observatory, Karl-Schwarzschild-Strasse 2, 85748 Garching bei München, Germany}

\author[0000-0002-5566-0634]{Tomoharu Oka}
\affiliation{Department of Physics, Faculy of Science and Technology, Keio University, 3-14-1 Hiyoshi, Kohoku-ku, Yokohama, Kanagawa 223-8522, Japan}

\author[0000-0002-3972-1978]{Jaime E. Pineda}
\affiliation{Max-Planck-Institut f\"ur extraterrestrische Physik, Giessenbachstrasse 1, D-85748 Garching, Germany}

\author[0000-0003-2133-4862]{Thushara G.S. Pillai}
\affiliation{Haystack Observatory, Massachusetts Institute of Technology, 99 Milsstone Road, Westford, MA 01886, USA}

\author[0000-0002-2887-5859]{V\'ictor M. Rivilla}
\affiliation{Centro de Astrobiología (CAB), CSIC-INTA, Carretera de Ajalvir km 4, Torrejón de Ardoz, 28850 Madrid, Spain}

\author[0000-0002-3078-9482]{\'Alvaro S\'anchez-Monge}
\affiliation{Institute of Space Sciences (ICE, CSIC), Campus UAB, Carrer de Can Magrans s/n, 08193, Bellaterra (Barcelona), Spain}
\affiliation{Institute of Space Studies of Catalonia (IEEC), 08860, Barcelona, Spain}

\author[0000-0002-3941-0360]{Miriam G. Santa-Maria}
\affiliation{Department of Astronomy, University of Florida, P.O. Box 112055, Gainesville, FL 32611}

\author{Howard A. Smith}
\affiliation{Center for Astrophysics $|$ Harvard \& Smithsonian, 60 Garden Street, Cambridge, MA 02138}

\author{Yoshiaki Sofue}
\affiliation{Institue of Astronomy, The University of Tokyo, Mitaka, Tokyo, 181-0015}

\author[0000-0001-6113-6241]{Mattia C. Sormani}
\affiliation{Department of Physics, University of Surrey, Guildford GU2 7XH, UK}
\affiliation{Universit{\`a} dell’Insubria, via Valleggio 11, 22100 Como, Italy}

\author[0000-0002-5445-5401]{Grant R.~Tremblay}
\affiliation{Center for Astrophysics $|$ Harvard \& Smithsonian, 60 Garden St., Cambridge, MA 02138, USA}

\author[0000-0002-4346-5858]{Gijs Vermari{\"e}n}
\affiliation{Leiden Observatory, Leiden University, P.O. Box 9513, 2300 RA Leiden, The Netherlands}

\author[0000-0001-8121-0234]{Alexey Vikhlinin}
\affiliation{Center for Astrophysics $|$ Harvard \& Smithsonian, 60 Garden St., Cambridge, MA 02138, USA}

\author[0000-0001-8504-8844]{Serena Viti}
\affiliation{Leiden Observatory, Leiden University, P.O. Box 9513, 2300 RA Leiden, The Netherlands}

\author{Dan Walker}

\author[0000-0002-9279-4041]{Q. Daniel Wang}
\affiliation{Department of Astronomy, University of Massachusetts, Amherst, MA 01003, USA}

\author[0000-0001-5950-1932]{Fengwei Xu}
\affiliation{Kavli Institute for Astronomy and Astrophysics, Peking University, Beijing 100871, People's Republic of China}
\affiliation{Department of Astronomy, School of Physics, Peking University, Beijing, 100871, People's Republic of China}

\author[0000-0003-2384-6589]{Qizhou Zhang}
\affiliation{Center for Astrophysics $\vert$ Harvard \& Smithsonian, 60 Garden Street, Cambridge, MA, 02138, USA}

\begin{abstract}
A compact source, G0.02467-0.0727, was detected in ALMA \threemm observations in continuum and very broad line emission.
The continuum emission has a spectral index $\alpha\approx3.3$, suggesting that the emission is from dust.
The line emission is detected in several transitions of CS, SO, and SO$_2$ and exhibits a line width FWHM $\approx160$ \kms.
The line profile appears Gaussian.
The emission is weakly spatially resolved, coming from an area on the sky $\lesssim1"$ in diameter ($\lesssim10^4$ au at the distance of the Galactic Center; GC).
The centroid velocity is $v_{LSR}\approx40$-$50$ \kms, which is consistent with a location in the Galactic Center.
With multiple SO lines detected, and assuming local thermodynamic equilibrium (LTE) conditions, $T_\mathrm{LTE} = 13$ K, which is colder than seen in typical GC clouds, though we cannot rule out low-density, subthermally excited, warmer gas.
Despite the high velocity dispersion, no emission is observed from SiO, suggesting that there are no strong ($\gtrsim10~\kms$) shocks in the molecular gas.
There are no detections at other wavelengths, including X-ray, infrared, and radio.

We consider several explanations for the Millimeter Ultra-Broad Line Object (MUBLO), including protostellar outflow, explosive outflow, collapsing cloud, evolved star, stellar merger, high-velocity compact cloud, intermediate mass black hole, and background galaxy.
Most of these conceptual models are either inconsistent with the data or do not fully explain it.
The MUBLO is, at present, an observationally unique object.

\end{abstract}
\section{Introduction}

The center of our Galaxy contains billions of stars, tens of millions of solar masses of gas, a supermassive black hole, a tenth of our Galaxy's ongoing star formation, and an extensive graveyard of stellar remnants \citep[e.g.,][]{Morris1996,Henshaw2023}.
It is therefore the likeliest place to find new classes of objects.
We present one such object in this work.

The following will describe observations (\S \ref{sec:observations}), measurements (\S \ref{sec:measurements}) of spectral (\S \ref{sec:spectralmeasurements}), spatial (\S \ref{sec:spatialmeasurements}), spatio-spectral (\S \ref{sec:spatio-spectral}), and continuum (\S \ref{sec:continuum}) data followed by analysis and modeling (\S \ref{sec:analysis}) of the excitation conditions (\S \ref{sec:LTE}, \ref{sec:nonlte}), chemistry (\S \ref{sec:chemistry}), and dust (\S \ref{sec:weirddust}).
We then discuss the location of the source (\S \ref{sec:where}), both along the line-of-sight (\S \ref{sec:los}) and on the sky (\S \ref{sec:spatiallocation}).
We engage in extended discussion of the similarities and differences between this object and other classes of objects (\S \ref{sec:WhatIsIt}) before concluding that we do not know exactly what this object is (\S \ref{sec:conclusion}).
Several appendices present additional spectra (\S \ref{appendix:totalpowerspectra}), detailed chemical models (\S \ref{appendix:chemistry}), and RADEX non-LTE models (\S \ref{appendix:radex}).

\section{Observations}
\label{sec:observations}
The ACES \rr{(ALMA CMZ Exploration Survey)} large program (2021.1.00172.L; PI Longmore) observed the Central Molecular Zone (CMZ) with ALMA \rr{(the Atacama Large Millimeter/Submillimeter Array)} in Band 3.
In brief, these data cover six windows: two medium-width covering 86--86.5 and 86.7--87.1~GHz, two broad covering 97.66--99.54 and 99.56--101.44~GHz, and two narrow windows covering 60~MHz centered on HNCO\,4--3 ($\nu_\mathrm{rest}=87.925238$~GHz) and HCO$^{+}$ ($\nu_\mathrm{rest}=89.18852$~GHz).
The latter two in particular were shifted to try to cover the full range of velocities of CMZ clouds, since their full bandwidth is only $\sim200$~\kms.
\rr{The ACES project covers the whole molecular component of the Galactic center, spanning roughly $-0.6\deg<\ell<0.9\deg$ and $-0.3\deg<b<0.2\deg$ with a total area 1200 square arcminutes, though in this work we focus only on the few arcsecond region around the MUBLO.}
Details of the observational setup are given in Table \ref{tab:data_props}.

The measurement sets were produced by the ALMA pipeline using CASA 6.4.1.12 pipeline 2022.2.0.64; these data were retrieved from the ALMA archive and restored on disk.
The data were imaged using CASA 6.4.3-2, adopting the same parameters as used in the original ALMA-delivered pipeline products, but with modifications as needed to fix bad cleans (specifically, iterative clean runs that diverged and produced spurious signals), to image windows that were left un-imaged because of size mitigation (the two broad-band windows were often excluded), or to image those said windows with full spectral resolution for the same reason.
The continuum data were imaged using the default parameters from the ALMA pipeline, including continuum identification in the UV domain from the ALMA pipeline's \texttt{findcont} task.
\rr{We have combined the ACES 3 mm data with MUSTANG images from \citet{Ginsburg2020} for display purposes in several figures, but all measurements given below are from the ALMA data alone.}
The full ACES data are still being processed, so as of this publication, we do not yet have a complete census of the broader context.

In this paper, we focus only on field \texttt{aa}, with MOUS \rr{(member observation unit set)} ID \texttt{A001\_X15a0\_X13c}.
During the quality assessment process for the ACES data reduction, we discovered an object with surprisingly large line width in several spectral lines.
We label this a Millimeter Ultra Broad Line Object, or MUBLO, since we do not know its nature beyond its observational properties. 

To verify that this feature was not an image artifact (though there was no particular reason to suspect it was), we searched the ALMA archive for overlap with this object.
Two programs observing the 50~\kms cloud\rr{, a molecular cloud centered at roughly $\ell=0.02\deg$, $b=-0.08\deg$ $v_{LSR}=50$ \kms \citep{Tsuboi2009},} covered this source.
Project 2012.1.00080.S (PI: Tsuboi) in Band 3, which overlaps with the ACES spectral coverage, and 2017.1.01185.S (PI: Mills) in Band 7 with only the 7m array, both \rr{cover the MUBLO}, though it is at the edge of the field in the latter and subject to high noise.
We re-imaged the archival data from both programs using the ACES pipeline.
We note that, while images and cubes were obtained from the 2012.1.00080.S project, and the results below are reasonable, there are some artifacts that persist in that data set that lead us to assign lower credence to differences between those and the ACES data.
The problems with the 2012 data set are severe enough that we choose not to show any of the images and strongly caution against interpreting the measurements of these data based only on their statistical errors.
Nevertheless, the spectral line measurements from the 2012 data show a few additional detections \rr{(see \S \ref{sec:spectralmeasurements})}, and the consistent detections of overlapping lines gives us high confidence that the ACES detections are not spurious.

The relevant observational parameters, including uncertainties and beam sizes, are given in Table~\ref{tab:data_props}.

\begin{table*}[htp]
\centering
\caption{Data Properties}
\begin{tabular}{ccccccccc}
\label{tab:data_props}
Data Type & Observation Date & Major & Minor & PA & RMS & Jy/K & $\nu_{min}$ & $\nu_{max}$ \\
 &  & $\mathrm{{}^{\prime\prime}}$ & $\mathrm{{}^{\prime\prime}}$ & $\mathrm{{}^{\circ}}$ & $\mathrm{mJy\,beam^{-1}}$ &  & $\mathrm{GHz}$ & $\mathrm{GHz}$ \\
\hline
Continuum&&&&&&&&\\
(2012.1.00080.S) & 2013-05-31T09:52:50.256001 & 2.359 & 1.397 & 86.991 & 0.4 & 44.0 & 85.2491 & 98.4496 \\
Continuum&&&&&&&&\\
(2017.1.01185.S) & 2018-09-27T21:30:24.144000 & 4.337 & 2.519 & 88.344 & 8.8 & 0.9 & 342.2384 & 357.9871 \\
Continuum spw33+35&&&&&&&&\\
(2021.1.00172.L) & 2022-09-15T00:26:52.128000 & 1.538 & 1.306 & -59.907 & 0.2 & 61.4 & 97.6648 & 101.4321 \\
Continuum spw25+27&&&&&&&&\\
(2021.1.00172.L) & 2022-09-15T00:26:52.128000 & 1.837 & 1.468 & -70.832 & 0.2 & 60.5 & 85.9664 & 87.1328 \\
spw25 & 2022-09-16T00:31:34.272000 & 1.926 & 1.573 & -71.850 & 3.1 & 54.3 & 85.9664 & 86.4328 \\
spw27 & 2022-09-16T00:31:34.272000 & 1.945 & 1.566 & -72.502 & 3.2 & 53.1 & 86.6665 & 87.1328 \\
spw33 & 2022-09-15T00:26:52.128000 & 1.626 & 1.440 & -52.088 & 3.4 & 53.7 & 97.6648 & 99.5394 \\
spw35 & 2022-09-16T00:31:34.272000 & 1.570 & 1.293 & -67.596 & 2.6 & 59.6 & 99.5619 & 101.4321 \\
\hline
\end{tabular}
\par

The data types labeled spw\texttt{nn} (spw is an abbreviation of spectral window) are from the ACES 2021.1.00172.L data.
The 2017.1.01185.S data are from the 7m array at ALMA, while the other data sets
are from the 12m array.

\end{table*}

\section{Measurements}
\label{sec:measurements}

We report measurements of the spectral lines (\S \ref{sec:spectralmeasurements}), the spatial location of both the lines and continuum (\S \ref{sec:spatialmeasurements}), the spatial-spectral structure (\S \ref{sec:spatio-spectral}), and the continuum (\S \ref{sec:continuum}) in this section.

\subsection{Spectral}
\label{sec:spectralmeasurements}

\begin{figure}[!ht]
    \includegraphics[width=0.49\textwidth]{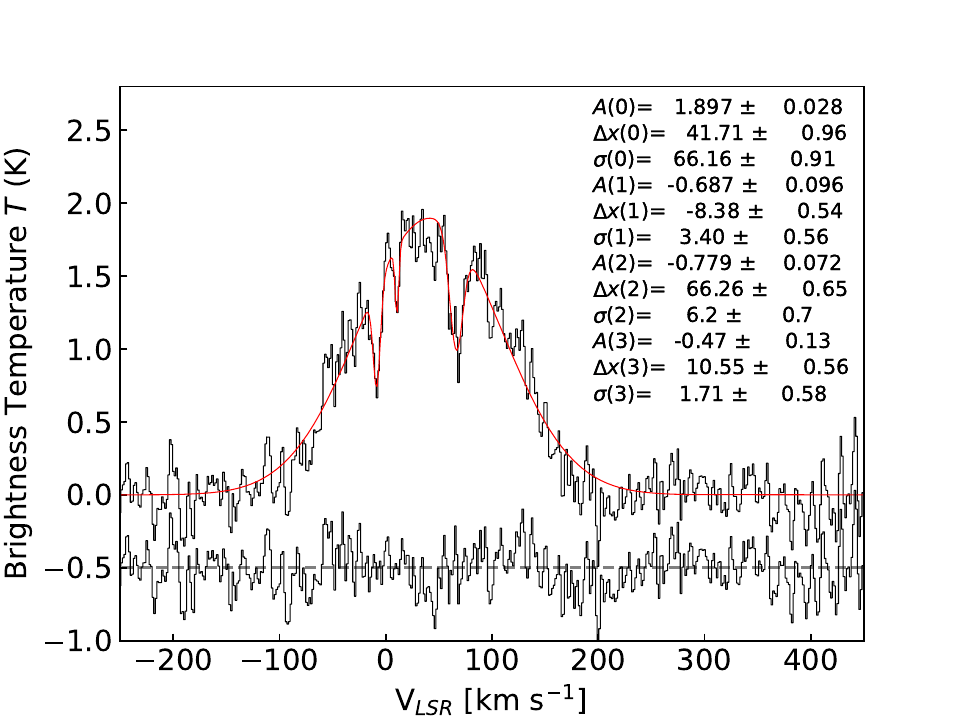}
    \includegraphics[width=0.49\textwidth]{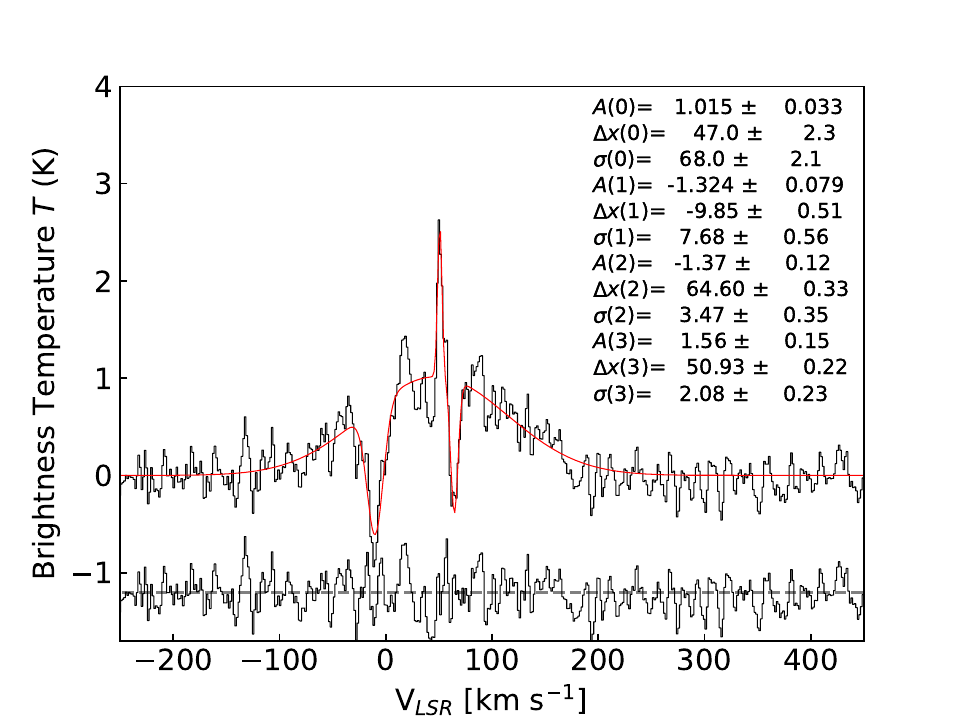}
    \caption{The SO 2(3)-1(2) [left] and CS 2-1 [right] spectra \rr{from the ACES data} with best-fit models, including narrow absorption and emission features, overlaid.
    The bottom spectrum shows the fit residual, with the dashed line indicating the zero level.
    \rr{The legend shows the best-fit parameters for the 1D Gaussian functions fit, $A$ is amplitude in K, $\Delta x$ is velocity in \kms, and $\sigma$ is the width in \kms.
    }
    }
    \label{fig:fittedspectra}
\end{figure}

We detect lines of CS, SO, and SO$_2$ listed in Table \ref{tab:spectral_measurements}.
We fitted Gaussian line profiles to the detected lines, and for CS 2-1 and SO 2(3)-1(2), we included foreground absorption components in our models (Figure~\ref{fig:fittedspectra}).
The line widths are $\sigma=60$--70~\kms (FWHM $\sim160$~\kms).

\begin{figure}[!ht]
    \centering
    \includegraphics[width=0.49\textwidth]{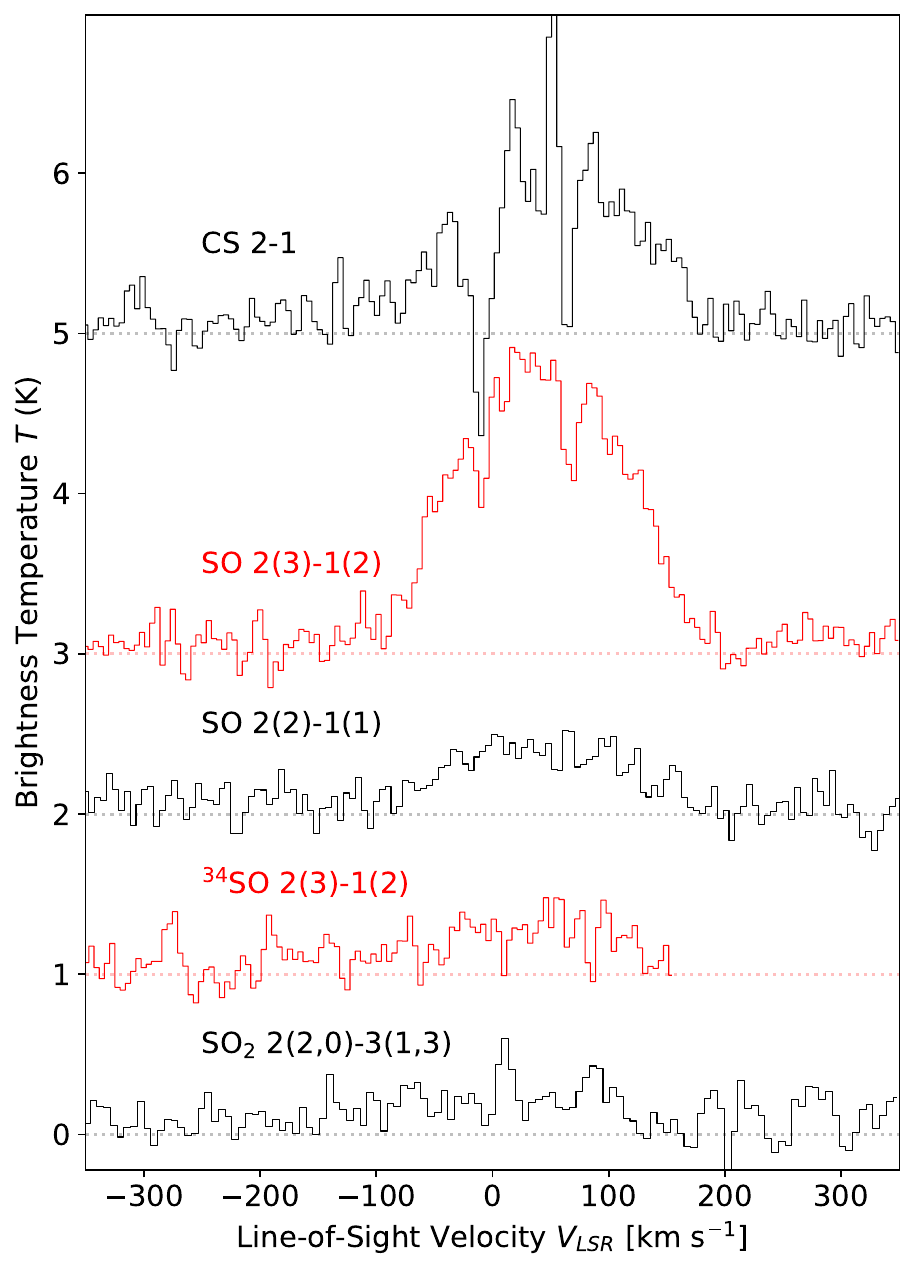}
    \includegraphics[width=0.49\textwidth]{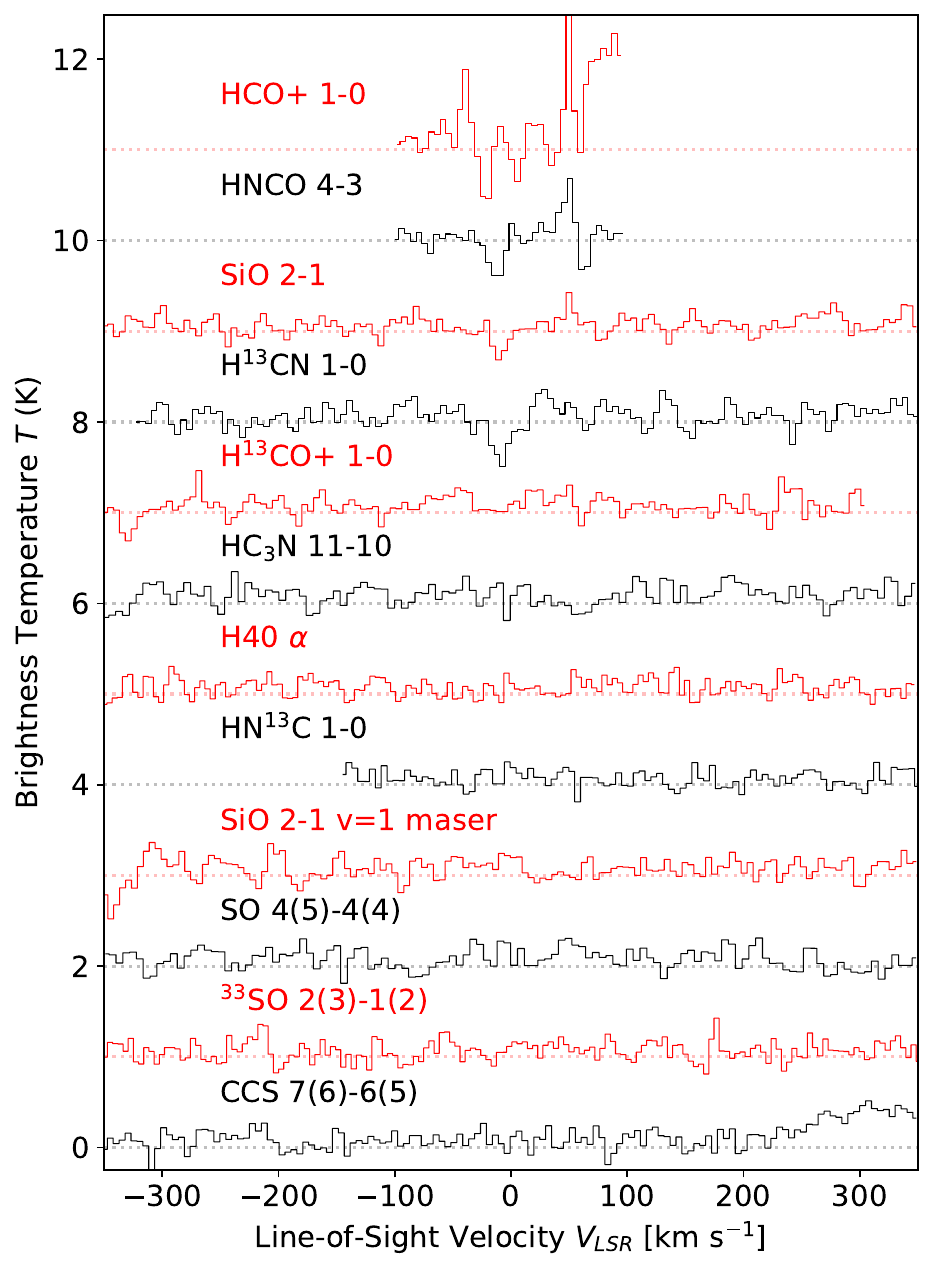}
    \caption{Spectra of \rr{detected lines} in the ACES spectral coverage (left) and \rr{relevant nondetections} (right).
    Only CS, SO, and SO$_2$ \rr{and their isotopologues} are detected.
    Appendix figure \ref{fig:coarse_spectra_withTP} shows the same data with the Total Power spectra, which covers larger physical scales ($\sim2$ pc), overlaid.
    }
    \label{fig:coarse_spectra}
\end{figure}

\begin{figure}[!ht]
    \includegraphics[width=0.49\textwidth]{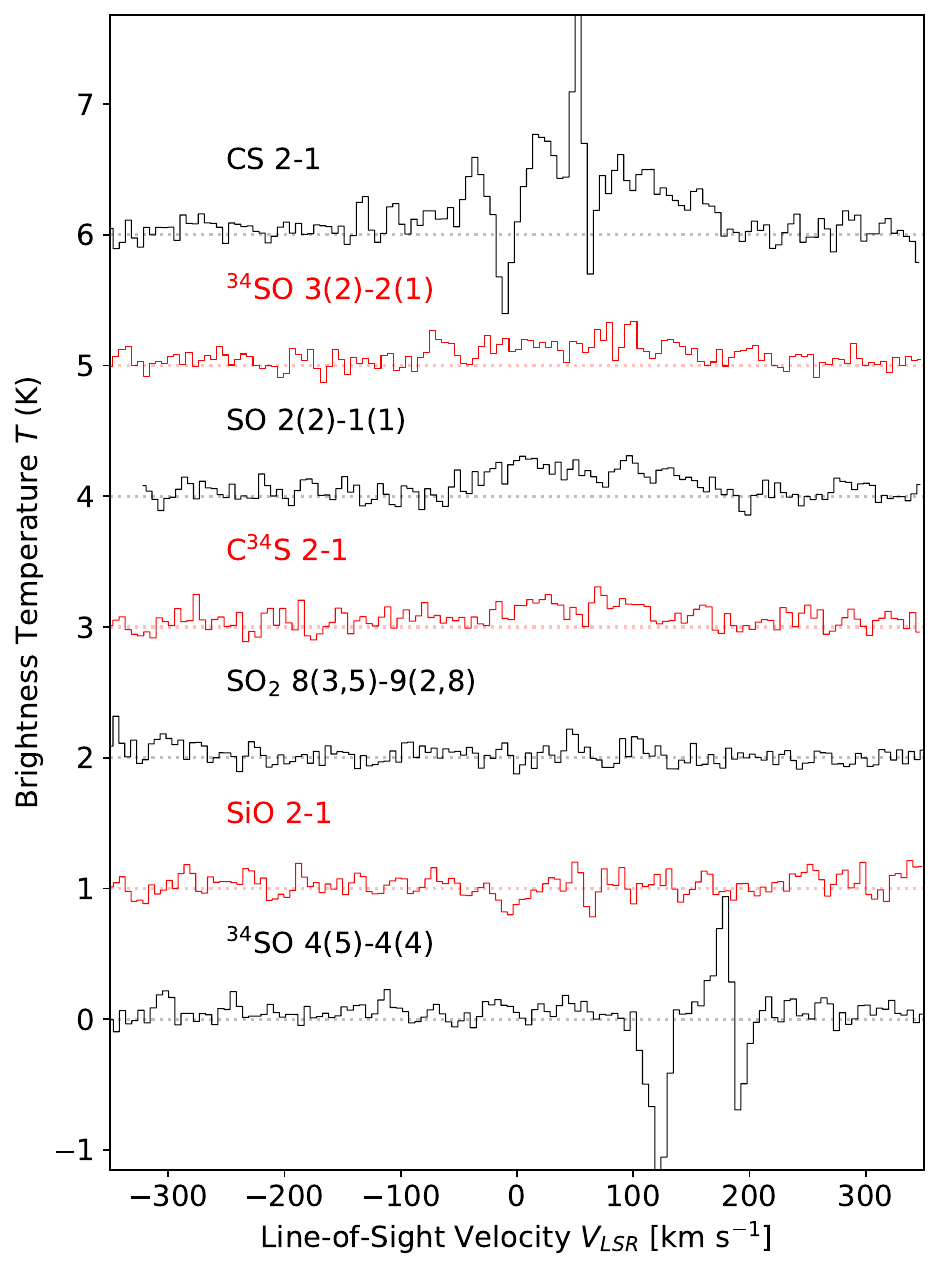}
    \includegraphics[width=0.49\textwidth]{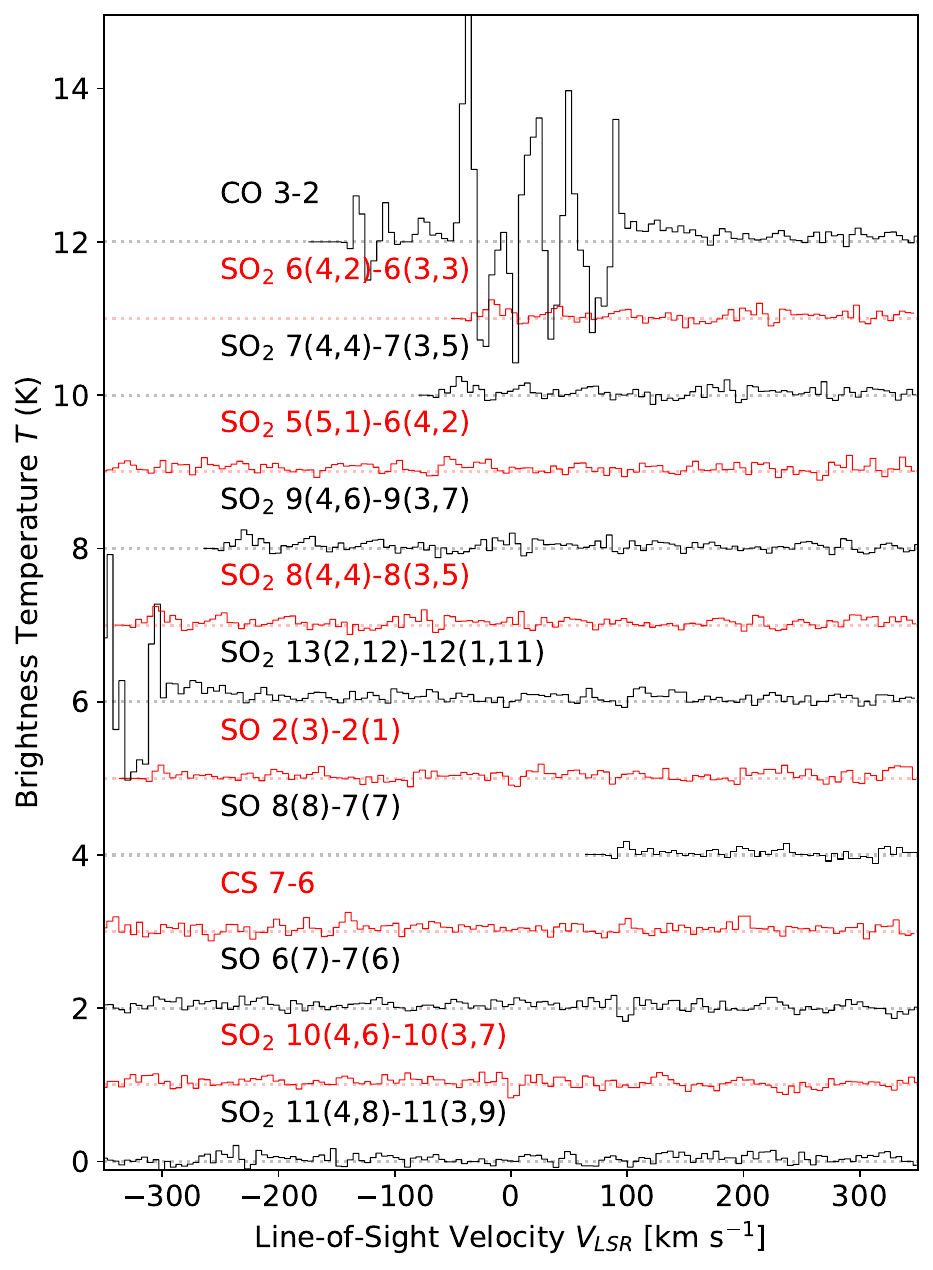}
    \caption{Spectra from 2012.1.00080.S in B3 (left) and 2017.1.01185.S in B7 (right).
    These show (marginal) detections of $^{34}$SO 2(3)-1(2) and $^{34}$CS 2-1 in B3 while confirming the clear detections of CS 2-1 and SO 2(2)-1(1). 
    There are no detections in B7.
    \rr{The features in the CO 3-2 spectrum (also seen at the edge of SO$_2$ 13(2,12)-12(1,11)) are from line-of-sight, extended features that are poorly reconstructed in the image cube. }
    }
    \label{fig:other_spectra}
\end{figure}

Figures~\ref{fig:coarse_spectra} and \ref{fig:other_spectra} show spectra of many candidate lines smoothed to 5~\kms resolution.
These spectra are extracted from the peak emission location, i.e., ICRS \rr{equatorial} coordinates 17h45m57.75s $-28\deg57'10.77"$.

\begin{table*}[htp]
\centering
\caption{Spectral Line Measurements}
\begin{tabular}{cccccccccc}
\label{tab:spectral_measurements}
Species & Rest Freq & Amp & $\sigma_{A}$ & Center & $\sigma_{\nu,cen}$ & $v_{cen}$ & $\sigma_{v,cen}$ & FWHM & $\sigma_{FWHM}$ \\
 & $\mathrm{GHz}$ & $\mathrm{K}$ & $\mathrm{K}$ & $\mathrm{GHz}$ & $\mathrm{GHz}$ & $\mathrm{km\,s^{-1}}$ & $\mathrm{km\,s^{-1}}$ & $\mathrm{km\,s^{-1}}$ & $\mathrm{km\,s^{-1}}$ \\
\hline
CS 2-1 & 97.980950 & 1.032 & 0.044 & 97.9656 & 0.0010 & 47.1 & 3.1 & 164.5 & 6.7 \\
SO 2(3)-1(2) & 99.299870 & 1.748 & 0.025 & 99.2866 & 0.0004 & 40.1 & 1.1 & 161.2 & 2.6 \\
$^{34}$SO 2(3)-1(2) & 97.715317 & 0.216 & 0.024 & 97.7039 & 0.0026 & 34.9 & 8.0 & 146.8 & 19.7 \\
SO 2(2)-1(1) & 86.093950 & 0.383 & 0.026 & 86.0829 & 0.0015 & 38.5 & 5.3 & 159.6 & 12.6 \\
SO$_2$ 2(2,0)-3(1,3) & 100.878105 & 0.172 & 0.029 & 100.8748 & 0.0048 & 9.9 & 14.3 & 173.9 & 33.6 \\
\hline \\
2012\\
\hline
C$^{34}$S 2-1  & 96.412951 & 0.139 & 0.021 & 96.3966 & 0.0034 & 50.8 & 10.6 & 139.8 & 24.9 \\
CS 2-1  & 97.980950 & 0.538 & 0.027 & 97.9676 & 0.0011 & 40.7 & 3.5 & 158.6 & 7.7 \\
$^{34}$SO 2(3)-1(2)  & 97.715317 & 0.129 & 0.015 & 97.6974 & 0.0031 & 54.8 & 9.4 & 168.8 & 22.1 \\
\hline
\end{tabular}
\par
The columns with $\sigma$ values give the uncertainty on their neighbors (e.g., $\sigma_A$ is the uncertainty on the amplitude).
All values are from Gaussian fits to the line profile, so the widths are the Gaussian width parameter, not the FWHM.

\end{table*}

\subsection{Spatial}
\label{sec:spatialmeasurements}
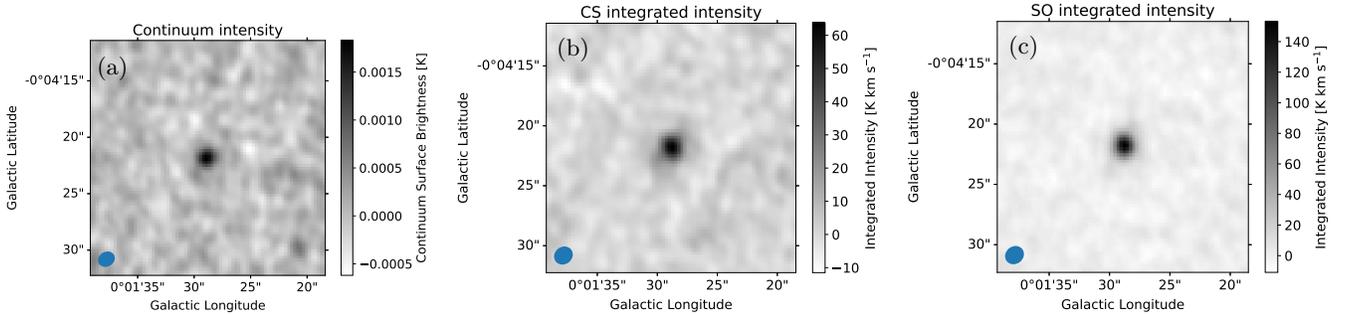
\begin{figure}
    \centering
    \begin{tikzpicture}
        \node[anchor=north west,inner sep=0pt] at (0,0){
        \includegraphics[width=0.32\textwidth]{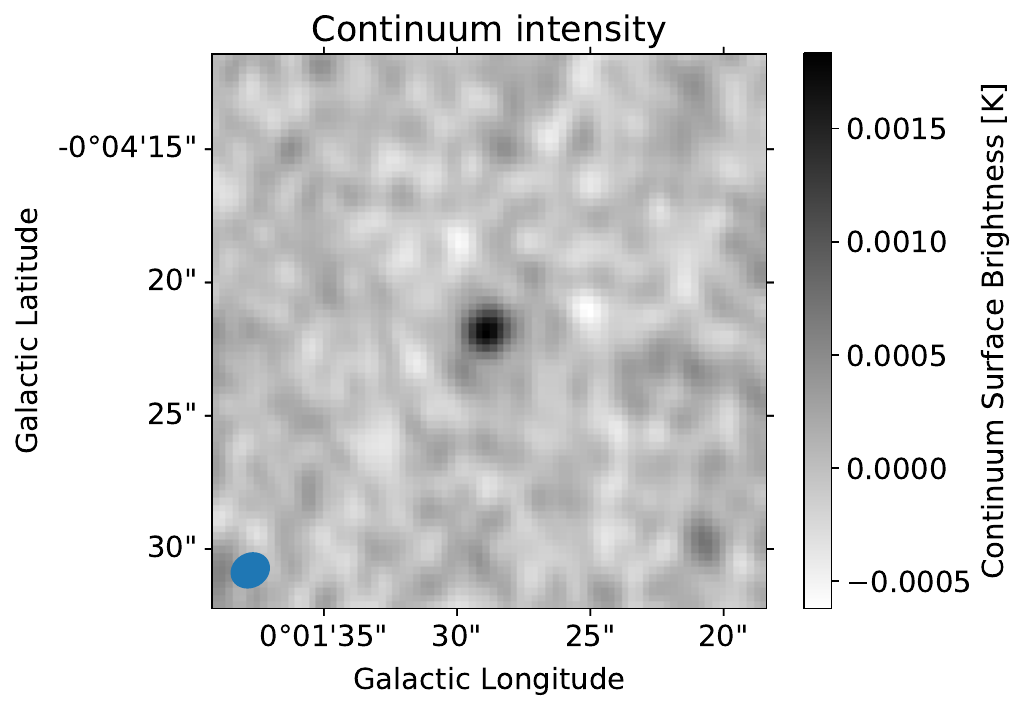}
        };
        \node[] at (11ex,-5ex) {(a)};
    \end{tikzpicture}
    \begin{tikzpicture}
        \node[anchor=north west,inner sep=0pt] at (0,0){
        \includegraphics[width=0.32\textwidth]{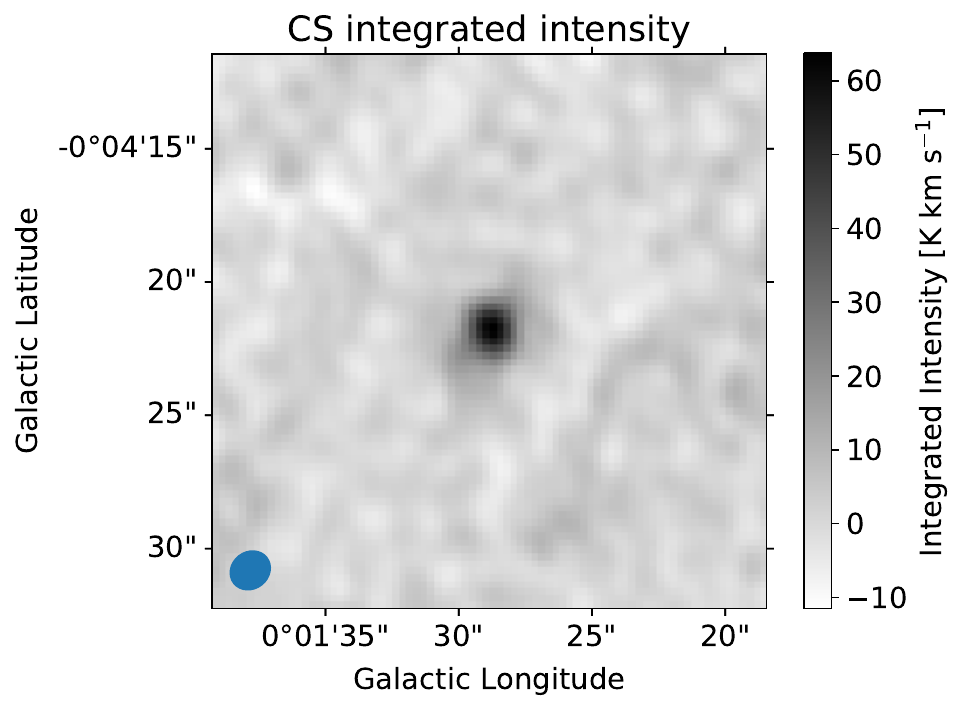}
        };
        \node[] at (12ex,-5ex) {(b)};
    \end{tikzpicture}
    \begin{tikzpicture}
        \node[anchor=north west,inner sep=0pt] at (0,0){
        \includegraphics[width=0.32\textwidth]{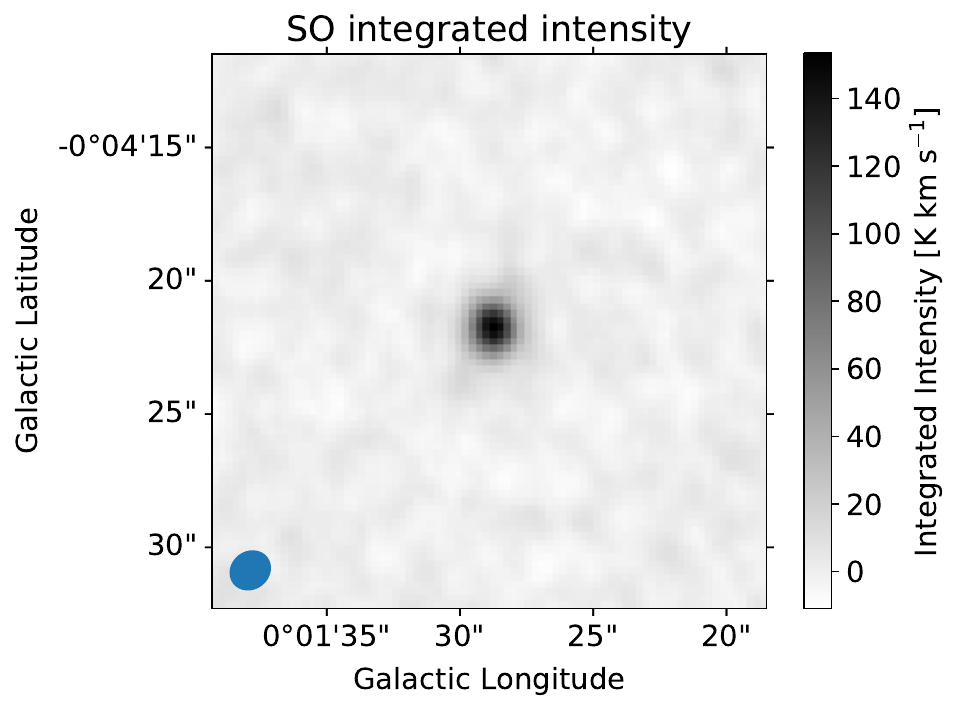}
        };
        \node[] at (12ex,-5ex) {(c)};
    \end{tikzpicture}
    
    \caption{Images of the source at three wavelengths: ACES \threemm Continuum (a), CS 2-1 moment 0 (b), SO 2(3)-1(2) moment 0 (c).  The moment 0 (integrated intensity) images exclude frequencies at which absorption is seen in the line profile or significant extended structure is detected around the source.  Moment 0 maps showing those velocities, which give a sense of the possible host environment, are shown in Figure \ref{fig:sourceoncloud}.}
    \label{fig:continuumandmom0}
\end{figure}

We fitted 2D Gaussian profiles to the continuum and integrated intensity images in all three data sets.
The fit parameters from both the continuum and spectral line fits are in Table~\ref{tab:spatial_measurements}.
\rr{The uncertainties in this table give only the statistical errors, but all of the spatial measurements are likely affected by systematic errors that are larger; we expect the dominant systematic uncertainty to be from separation between the structured background brightness and the compact Gaussian.
While the statistical errors appear to show significant offsets between the different measurements, we regard these as unlikely to be real given the systematic uncertainty.
}
Figure~\ref{fig:continuumandmom0} shows the \threemm continuum image from the ACES data alongside the integrated intensity (moment-0) images of CS 2-1 and SO 2(3)-1(2).

\begin{table*}[htp]
\centering \small
\caption{Spatial Measurements}
\begin{tabular}{ccccclll}
\label{tab:spatial_measurements}
Image & Amp & Amp[mJy] & RA off & Dec. off & Major & Minor & PA \\
 & $\mathrm{K}$ & $\mathrm{mJy}$ & $\mathrm{marcsec}$ & $\mathrm{marcsec}$ & $\mathrm{{}^{\prime\prime}}$ & $\mathrm{{}^{\prime\prime}}$ & $\mathrm{{}^{\circ}}$ \\
\hline
SO 2(3)-1(2) m0 & 148.3 (1.1) & 2795.0 (21.0) & -149 (5) & 4 (5) & 1.549 (0.012) & 1.816 (0.014) & 205.1 (1.9) \\
 &  &  &  &  & 0.85 & 0.51 & -81.16 \\
CS 2-1 m0 & 55.65 (0.76) & 1021.0 (14.0) & -22 (14) & -6 (11) & 2.555 (0.035) & 1.928 (0.026) & 86.0 (1.9) \\
 &  &  &  &  & 2.04 & 1.17 & 80.22 \\
CS 2-1 2012 m0 & 39.36 (0.46) & 1459.0 (17.0) & 411 (21) & 96 (16) & 4.34 (0.051) & 3.299 (0.039) & 95.9 (1.7) \\
 &  &  &  &  & 3.38 & 2.80 & -78.42 \\
Continuum spw33+35 & 0.1036 (0.0029) & 1.77 (0.05) & 20 (20) & 62 (18) & 1.73 (0.049) & 1.499 (0.043) & 85.3 (8.0) \\
 &  &  &  &  & 0.99 & 0.43 & 60.15 \\
Continuum spw25+27 & 0.1077 (0.0022) & 1.758 (0.035) & -55 (16) & 149 (15) & 1.971 (0.04) & 1.783 (0.036) & 62.4 (8.2) \\
 &  &  &  &  & 1.22 & 0.21 & 33.67 \\
Continuum 2012 cont & 0.0906 (0.0031) & 1.479 (0.05) & -34 (27) & 304 (18) & 1.958 (0.066) & 1.245 (0.042) & 436.3 (2.9) \\
 &  &  &  &  & - & - & - \\
Continuum B7 & 0.0896 (0.0037) & 98.2 (4.0) & 371 (53) & 569 (40) & 1.3 (0.13) & 0.993 (0.096) & 88.3 (6.1) \\
 &  &  &  &  & - & - & - \\
\hline
\end{tabular}
\par
The Amp and Amp[mJy] values are in K \kms and Jy/beam \kms for the spectral lines.
The RA off and Dec. off columns give offsets from the coordinate 17h45m57.753s -28d57m10.769s.
Under the Major, Minor, and PA columns, the alternating rows with no errorbars give the beam-deconvolved sizes.

\end{table*}

\subsection{Spatio-Spectral}
\label{sec:spatio-spectral}
The integrated and peak intensity maps of the detected emission lines are spatially weakly resolved.
The deconvolved size is $\sim1\arcsec\times0.5\arcsec$ (see Table \ref{tab:spatial_measurements}).

\begin{figure}
    \centering
    \includegraphics[width=0.45\textwidth]{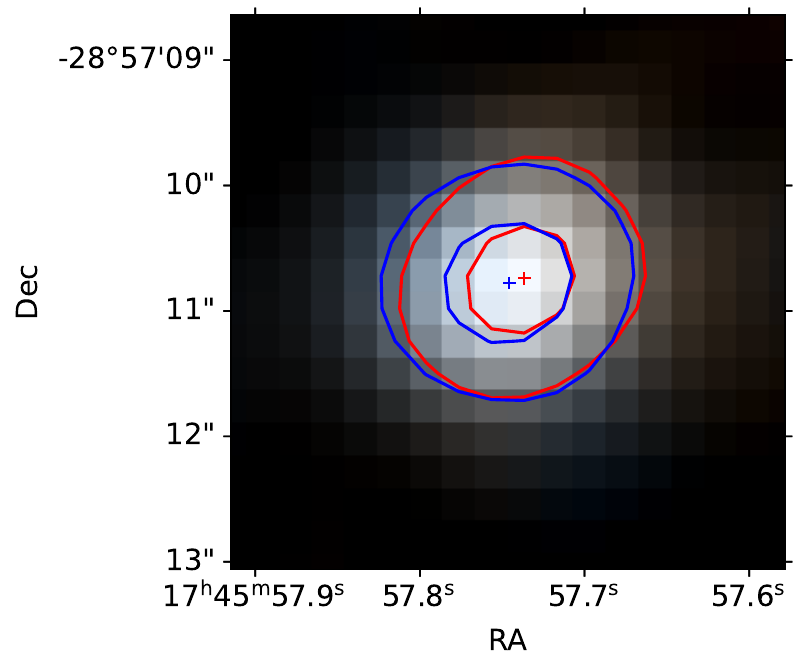}
    \includegraphics[width=0.45\textwidth]{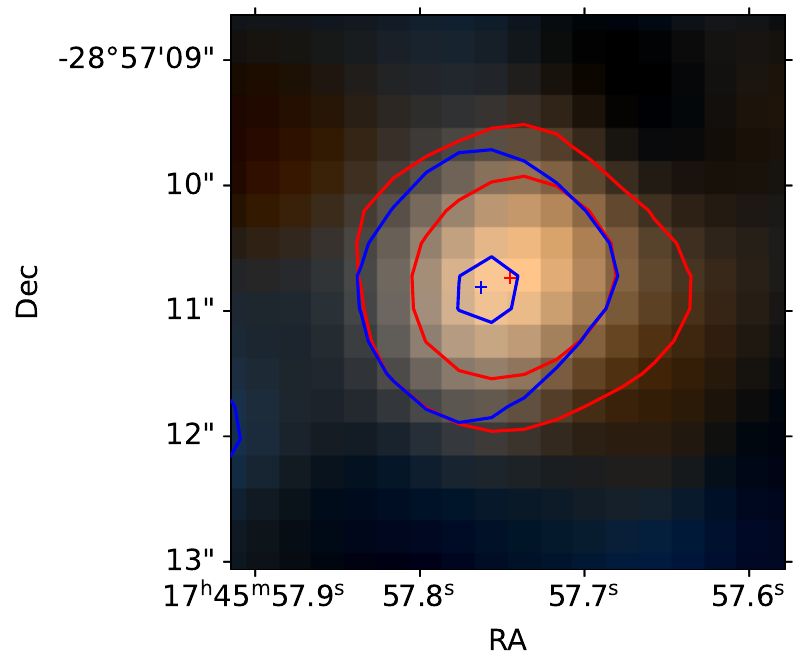}
    \caption{(left) Moment map made from integrating the red (64 to 152 \kms, weighted 104 \kms) and blue (--69 to +35 \kms, weighted --8 \kms) sides of the SO 2(3)-1(2) line profile.
    \rr{Contours are shown at 10 and 20 $\sigma$.}
    The plus symbols show the 2D Gaussian fit centroids for each moment map.
    They are separated by 0.5$\pm0.06$ pixels (0.23$\pm0.03$\arcsec, or $2000\pm240$ au).
    Both the image and contours show the same data.
    (right) Same for CS, but with more limited velocity ranges.  The centroid separation has the same general direction but is less significant, with a measured offset 0.46$\pm0.12$\arcsec - consistent at the $2\sigma$ level.  The CS integral was taken from -120 to -20 \kms on the blue side to avoid contamination from Galactic Center clouds, exaggerating the offset and reducing the signal-to-noise ratio.  The red side is integrated from 75 to 200 \kms.
    }
    \label{fig:SOredblue}
\end{figure}

There is a weak sign of a spatial velocity gradient.
We fitted 2D Gaussian profiles to each channel in the SO 2(3)-1(2) cube, but found that the fits were too unreliable at the modest S/N in each channel.
We therefore collapsed the red (64 to 152~\kms) and blue ($-68$ to 34~\kms) sides of the cube into moment-0 images, excluding an extended cloud that created narrow absorption and emission features around 45~\kms.
Figure~\ref{fig:SOredblue} shows the result of this fitting.
We find that the red and blue sides of the spectrum are spatially separated at $\sim10\sigma$ significance \rr{(and unlike the measurements in \ref{sec:spatialmeasurements}, the systematic uncertainties in the red and blue measurements are expected to cancel out)}.
The separation is 2000$\pm200$~au assuming $d=8$~kpc over a 112~\kms difference, resulting in a gradient of 11,000~\kms~pc$^{-1}$.
Repeating the same measurement with CS 2-1 shows a gradient in the same direction, but with a much larger amplitude and lower signal-to-noise ratio; the CS measurement was more affected by contamination from extended CMZ clouds.

\subsection{Continuum}
\label{sec:continuum}
We detect continuum emission in bands 3 and 7.
Assuming the continuum comes from a single point source, which is consistent with the measurements in Table \ref{tab:spatial_measurements}, the spectral index is $\alpha=3.25\pm0.06$ (statistical) $\pm0.17$ (systematic, adopting 10\% calibration uncertainty),  indicating that the emission is coming from dust that is mostly optically thin.
Adopting a standard dust opacity for protostellar cores \citep[$\kappa_\mathrm{100 GHz} = 0.002$ cm$^{2}$ g$^{-1}$ extrapolated from][however, see \S \ref{sec:weirddust}]{Ossenkopf1994}, assuming T=20 K and gas\rr{-to-}dust \rr{mass ratio} 100 (so the mass in $\kappa$ is the total gas+dust mass), we estimate the mass from both frequencies:
\begin{equation}
    M=47~\msun \left(\frac{S_\mathrm{102 GHz}}{1.8 \mathrm{mJy}}\right) \left(\frac{T}{20\mathrm{~K}}\right)^{-1} \left(\frac{\kappa}{0.002\mathrm{~cm^{2}~ g^{-1}}}\right)^{-1} \left(\frac{D}{8 \mathrm{~kpc}}\right)^2,
\end{equation}
\begin{equation}
M=35~\msun \left(\frac{S_\mathrm{350 GHz}}{98 \mathrm{mJy}}\right) \left(\frac{T}{20\mathrm{~K}}\right)^{-1} \left(\frac{\kappa}{0.018\mathrm{~cm^{2}~ g^{-1}}}\right)^{-1} \left(\frac{D}{8 \mathrm{~kpc}}\right)^2.
\end{equation}
The millimeter dust opacity is low, $\tau<0.01$, at both frequencies.
The column density, $N(H_2)\sim4\times10^{23}$ \persc, corresponds to extinction $A_K\approx20$.
For the \rr{Band 3 (B3)} source size of $r\approx5000$ au, assuming spherical symmetry, the molecular number density is $n(\mathrm{H}_2)\sim10^7$ \percc.
Combining the dust mass with the CS and SO line widths, the energy in the gas is very large, $E_{kin} = 5\times10^{48}\mathrm{~erg} \left(\frac{M}{50 \msun}\right) \left(\frac{\sigma_v}{70 \kms}\right)^2$.

We caution that the dust properties assumed above might not be appropriate for all of the types of sources considered in the explanations for the MUBLO discussed in \S\ref{sec:WhatIsIt}.
Some aspects of non-standard dust properties are discussed in \S\ref{sec:weirddust}.

\begin{figure}[!ht]
    \centering
    \includegraphics[width=0.5\textwidth]{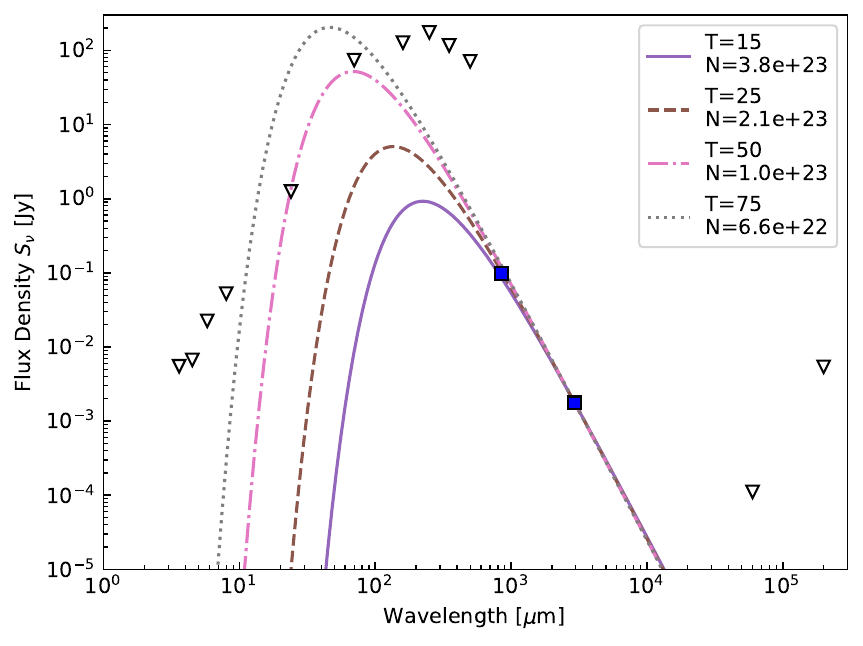}
    \caption{Spectral Energy Distribution of the MUBLO.
    The data at $\lambda \leq 24\um$ are from Spitzer, those from 70-500 \um are from Herschel, the millimeter-wavelength data - the detections - are from ALMA, and the longer-wavelength data are from the VLA \citep{Lu2019} and MEERKAT \citep{Heywood2022}.
    Triangles indicate upper limits.
    Four curves are overplotted showing modified blackbody models as described in \S \ref{sec:continuum}, with temperature indicated in Kelvin and column density in \persc.
    All models adopt dust opacity spectral index $\beta=1.5$.
    }
    \label{fig:SED}
\end{figure}

To further constrain the dust properties, we extract limits from multiwavelength data.
From the Spitzer \citep{Ramirez2008,Carey2009} and Herschel \citep{Traficante2011} data, we adopt the surface brightness at the position of the source as an upper limit, since in all wavelengths, there is significant extended emission.
Table \ref{tab:sed} gives the upper limits we determine at each wavelength.
Figure \ref{fig:SED} shows the SED with modified blackbody models overlaid.
The modified blackbodies are labeled with the adopted temperature in K and column density in \persc assuming $\beta=1.5$.
For a modified blackbody with a power law opacity function, $\kappa\propto\nu^\beta$, the allowed range of dust temperatures is $15 \lesssim T \lesssim 50$ K: the 850 \um~data point sets the temperature lower limit, while the 24 \um~and 70 \um~upper limits from Spitzer and Herschel data set the temperature upper limits (Figure \ref{fig:SED}).

\begin{table*}[htp]
\centering
\caption{Spectral Energy Distribution}
\begin{tabular}{cccc}
\label{tab:sed}
Wavelength & Surface Brightness & Beam Area & Flux \\
$\mathrm{\mu m}$ & $\mathrm{MJy\,sr^{-1}}$ & $\mathrm{sr}$ & $\mathrm{Jy}$ \\
\hline
3.6 & 51.613 & 1.065e-10 & 0.0055 \\
4.5 & 62.788 & 1.065e-10 & 0.00669 \\
5.8 & 211.374 & 1.065e-10 & 0.0225 \\
8.0 & 492.624 & 1.065e-10 & 0.0525 \\
24.0 & 1304.027 & 9.588e-10 & 1.25 \\
70.0 & 26726.094 & 2.764e-09 & 73.9 \\
160.0 & 26034.621 & 4.887e-09 & 127 \\
250.0 & 12025.196 & 1.451e-08 & 175 \\
350.0 & 4771.450 & 2.442e-08 & 117 \\
500.0 & 1484.628 & 4.794e-08 & 71.2 \\
850.0 & 299.154 & 3.283e-10 & 0.0982 \\
2939.1 & 18.810 & 9.412e-11 & 0.00177 \\
60000.0 & 1.481 & 7.484e-11 & 0.000111 \\
200000.0 & 12.638 & 4.261e-10 & 0.00539 \\
\hline
\end{tabular}
\par
Except for the ALMA measurements at 3 mm and 850 \um, these values are upper limits.
\end{table*}

\clearpage

\section{Analysis}
\label{sec:analysis}

In this section, we attempt to measure the gas temperature by modeling the SO lines under simple conditions (\S \ref{sec:LTE}).
We then allow for greater complexity in the excitation (\S \ref{sec:nonlte}) and chemistry (\S \ref{sec:chemistry}).
Finally, we explore the possibility that the dust has atypical properties (\S \ref{sec:weirddust}).

\subsection{LTE Modeling}
\label{sec:LTE}
\begin{figure}[!ht]
    \centering
    \includegraphics[width=0.5\textwidth]{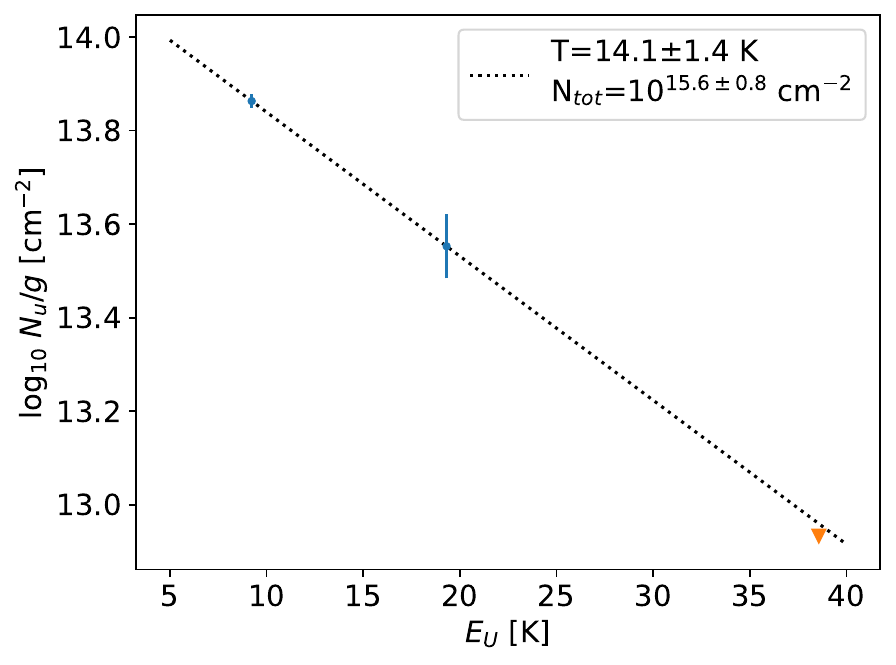}
    \caption{Rotation diagram showing the best fit for the SO lines.
    The orange point shows the upper limit from the 5(4)-4(4) line.
    }
    \label{fig:LTErotationdiagram}
\end{figure}

We detect two transitions of SO with different energy levels and obtain an upper limit on a third, allowing us to make a rotational diagram and fit the temperature and column density assuming local thermodynamic equilibrium (LTE) conditions, which is reasonable assuming the above density.
The temperature from the SO 2(3)-1(2) to 2(2)-1(1) line ratio is $T=14.1\pm 1.4$ K (Figure \ref{fig:LTErotationdiagram}; however, see \ref{sec:nonlte}).

\begin{figure}[!ht]
    \centering
        \begin{tikzpicture}
            \node[anchor=north west,inner sep=0pt] at (0,0){
            \includegraphics[width=0.49\textwidth]{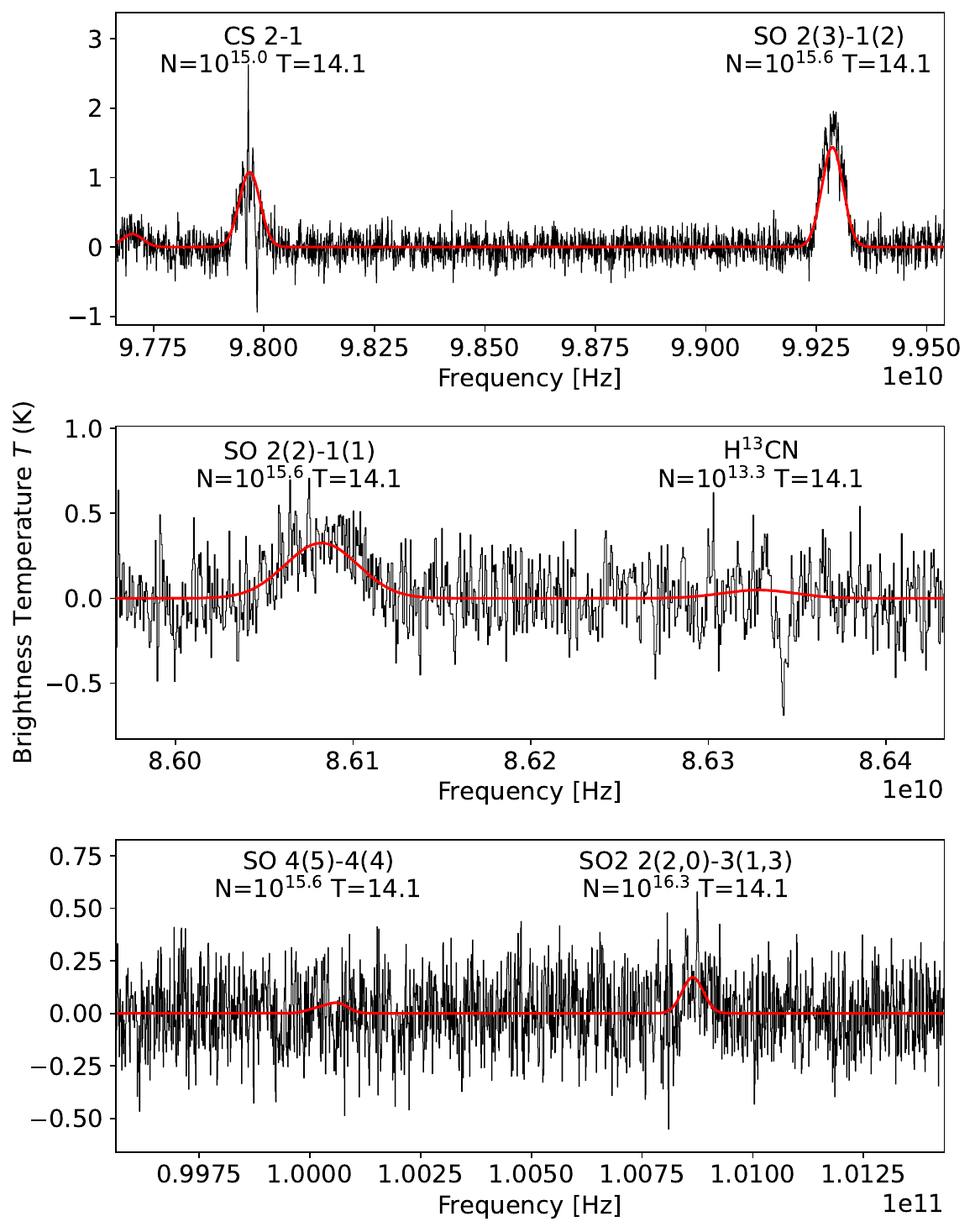}
            };
            \node[font=\bfseries\large] at (10ex,2ex) {(a)};
        \end{tikzpicture}
        \begin{tikzpicture}
            \node[anchor=north west,inner sep=0pt] at (0,0){
            \includegraphics[width=0.49\textwidth]{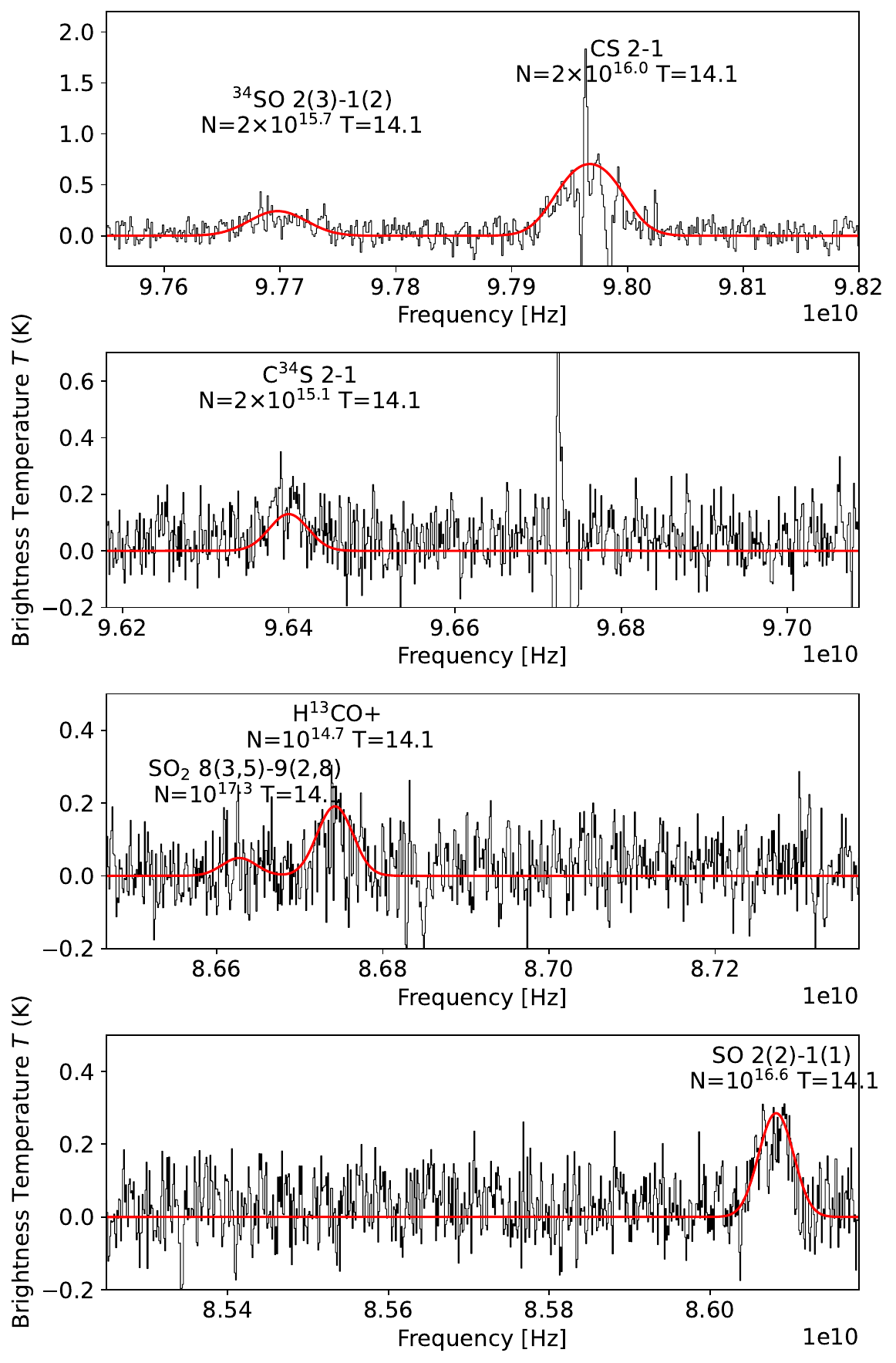}
            };
            \node[font=\bfseries\large] at (11ex,-4ex) {(b)};
        \end{tikzpicture}
    \caption{(a, left) The ACES spectrum. Two detections of SO, and one upper limit, let us put a constraint on the temperature and column density. 
    \rr{The red curve shows an LTE model spectrum with physical parameters shown in the text labels and velocity $v=41$ \kms, FWHM$=167$ \kms.}
    (b, right) The 2012.1.00080.S spectrum.
    Note that the apparent detection of H$^{13}$CO$^+$ is not from the MUBLO - while there is emission at this velocity, it is spatially diffuse, not associated with the MUBLO.
    }
    \label{fig:LTEmodel}
\end{figure}

We overplot a model emission spectrum based on the LTE fit on the spectral data.
Figure \ref{fig:LTEmodel}a shows the best-fit model for SO overlaid on the ACES data, and Figure \ref{fig:LTEmodel}b shows the model on the 2012 data.
In both figures, the CS and SO$_2$ column densities are scaled to fit the data while assuming the fitted SO rotational temperature.
Lines of C$^{34}$S and $^{34}$SO are also detected, giving a ratio $^{32}$S/$^{34}$S $\approx$ 8, consistent with some measurements in the Galactic center \citep{Yu2020} but inconsistent with others that find $^{32}$S/$^{34}$S$=16\pm4$ \citep{Humire2020} or $=19\pm2$ \citep{Yan2023}.
The column density is $\mathrm{N(SO)}=4\times10^{15}$~\persc.
Assuming the same temperature for the other molecules, we obtain column densities N(CS)=$10^{15}$ \persc, N(C$^{34}$S)=$1.2\times10^{14}$~\persc, N(SO$_2$)=$2\times10^{16}$~\persc, and N($^{34}$SO)=$5\times10^{14}$~\persc.
These estimates are dominated by systematic uncertainty in the excitation temperature, which may be more than an order of magnitude.
At these column densities, the peak line optical depth is $\tau \sim 0.1$, so the lines are still well-approximated as optically thin.
    
In these same figures (Fig. \ref{fig:LTEmodel}), we show model emission lines with the same centroid, linewidth, and excitation temperature for non-detected lines that we would expect to see in typical molecular and/or shocked gas: \hcopthirt, H$^{13}$CN, and SiO.
The models correspond to parts of the spectrum with no detection, so they give a rough upper limit on the column densities of these molecules.
\rr{For \hcopthirt, the figure appears to show a detection, but this emission comes from the diffuse medium, not the MUBLO.}

SO and CS appear to be $\gtrsim100\times$ more abundant than  H$^{13}$CO$^{+}$, H$^{13}$CN, HC$_3$N, and SiO.
If we adopt the dust-derived column density $N(\mathrm{H
}_2)\approx4\times10^{23}$ \persc, the abundances \rr{relative to H$_2$} are $X_\mathrm{SO}\approx10^{-8}$, $X_\mathrm{CS}\approx2\times10^{-9}$, $X_\mathrm{H^{13}CO^+} \sim X_\mathrm{H^{13}CN} < 5\times10^{-11}$, $X_\mathrm{SiO}<10^{-10}$, and $X_\mathrm{HC_3N} < 1\times10^{-10}$.

\subsection{Non-LTE conditions?}
\label{sec:nonlte}
If our mass measurement above is overestimated, the SO lines could be out of LTE, which could significantly change the above abundance and column density calculations.

\rr{Non-LTE models confirm the LTE column density estimate.}
We ran a grid of RADEX \rr{\citep{vanderTak2007}} models to test whether non-LTE conditions can match the data.
The most useful constraint on the non-LTE physical conditions comes from the intensity ratio of SO 2(3)-1(2)/2(2)-1(1), since this ratio should be mostly unaffected by the unknown filling factor of the emission.
Using this measured ratio, $R_{32}=4.6\pm0.3$, and the lower limit on the SO 2(3)-1(2) intensity $S_{32} > 1.75$ K from the filling factor $ff\leq1$, only a narrow range of parameter space is allowed in LTE models: $5\ee{15} < N(\mathrm{SO}) < 2\ee{16}$ \persc and $5< T < 13$~K.
RADEX one-zone models, adopting $dv=70$ \kms, give a wide range of solutions for different temperatures.
For example, for T=50 K, the H$_2$ density can be $10^{2.5} < n(\mathrm{H}_2) < 10^5$ \percc for $10^{16} > N(\mathrm{SO}) > 10^{15}$ \persc (see Appendix \ref{appendix:radex}).
Values \rr{of column density} $>0.5$ dex from the LTE model are not allowed even under non-LTE conditions, though the temperature is essentially unconstrained by the RADEX models.
However, if we incorporate our estimate of the H$_2$ number density based on the dust, the non-LTE models are ruled out: all of the high-temperature ($T\gtrsim20$ K) models require low densities ($n(\mathrm{H}_2)<10^{5.5}$ \percc).
Additionally, the Meudon PDR models that match the low HCN/CS upper limit and the CS/SO ratio require high density ($n(\mathrm{H}_2)\gtrsim10^7$ \percc); see \S \ref{sec:chemistry} below and Appendix~\ref{appendix:chemistry}.
We therefore disfavor the non-LTE, low-density model, but additional observations to further test this hypothesis by imaging other SO lines are straightforward and should be performed.

\subsection{Chemical Modeling}
\label{sec:chemistry}
We have run chemical models to search for physical parameter space consistent with the observations.
We ran both Meudon PDR \citep{LePetit2006} and UCLCHEM \citep{Holdship2017} models.

Using the time-dependent gas grain open-source chemical code UCLCHEM\footnote{\url{https://uclchem.github.io}} \citep{Holdship2017}, we ran models of a collapsing cloud that varied in final densities, UV irradiation, cosmic ray ionization rate, and temperatures.
The complete description of the models can be found in Appendix \ref{sec:uclchem}.
In brief, there is ample room in radiation field-temperature-density-cosmic ray ionization rate parameter space that produces high SO/SiO ratios (SO/SiO$>100$).
This ratio is produced by more models for longer time periods at higher densities ($n\gtrsim10^6$~\percc) and temperatures ($T\sim50$~K).
With the present data set, \rr{these chemical} models do not rule out any of the physical models considered below, but they point in constructive directions for what can be observed next to better understand the MUBLO.

We ran Meudon PDR\footnote{\url{https://pdr.obspm.fr/}} \citep{LePetit2006} models spanning a range of extinction, cosmic ray ionization rate, density, and UV field \rr{(see Appendix \ref{appendix:chemistry})}.
The Meudon model predicts line intensities in addition to abundances, so we compare to the predicted intensities for these models.
The observed line ratio SO 3(2)$-$2(1)\,/\,CS 2$-$1 can be reproduced at high density ($n_\mathrm{H}\sim10^7$~\percc) for a wide range of cosmic ray ionization rates (CRIR, $10^{-17} < \zeta_\mathrm{CR} < 10^{-15}$ \pers). 
At lower density, $n_\mathrm{H}\sim10^5$, the intensity ratio is at least an order of magnitude below what we observe.
The Meudon models therefore favor higher densities and support adopting LTE conditions for SO excitation modeling.

\subsection{Weird dust?}
\label{sec:weirddust}

Since we do not know that the dust is protostellar, we evaluate other possibilities.  
\citet{Bianchi2007} adopt a simple power law for dust in supernovae, which becomes $\kappa_\mathrm{100 GHz}=0.4\left(\frac{100\um}{3 \mathrm{mm}}\right)^{1.4}=0.0034$ cm$^2$ g$^{-1}$, a factor of two larger than we assumed; supernova dust would be only marginally different from our assumptions.
Following \citet{Kaminski2019}, who modeled dust in the circumstellar envelope of evolved star VY CMa, we extrapolate the \citet{Draine1984} opacities to be $\kappa_\mathrm{102 GHz}=0.00032$ and $\kappa_\mathrm{350 GHz}=0.0040$.
If the dust is like that in VY CMa, the mass is substantially ($\sim5\times$) larger than we reported.
\citet{Draine2006} gives a range of dust opacities from $0.0003<\kappa_\mathrm{102 GHz} < 0.03$~cm$^2$~g$^{-1}$, where the large end of this range corresponds to carbonaceous dust (pyrolized cellulose) that is too opaque to comprise a significant amount of the ISM.
If the dust is comprised primarily of carbon, the mass may be as much as $15\times$ smaller than we reported, a mere 2 \msun of gas, though this possibility is especially unlikely given the large column density of SO and SO$_2$ detected, which indicate that the medium is not especially carbon-rich.

In all of the above measurements, we have adopted a standard gas-to-dust ratio of 100.
If we were looking at a hydrogen-free object, comprised entirely of dust, the mass would be quite small, merely $\sim0.5$ \msun.
Such t{\"o}tally m{\"e}tal objects have been suggested to be possible to assemble in the turbulent ISM \citep{Hopkins2014}, but there is no immediate reason to expect it to exhibit extreme line widths.

Reservoirs with $\sim{}0.5\,\msun$ in dust alone have been inferred in some supernova remnants and evolved stars.
\citet{Kaminski2019} suggests that the envelope of VY~CMa might contain $0.5\,\msun$ of dust (see also \S \ref{sec:evolvedstars} below), but (as \citeauthor{Kaminski2019} points out) this value is uncertain because of the substantial optical depths involved in this calculation.
\citet{Chawner2019} infer dust reservoirs of 0.3--$0.5\,\msun$ in pulsar wind nebulae based on Herschel data.
Given stellar masses $\ll10^2\,\msun$ that produced these quantities of dust, there must be gas-to-dust mass ratios $\ll10^2$, higher dust opacities than assumed in the respective analysis, or both.
Such work demonstrates that, under certain conditions, continuum emission at an intensity seen in the MUBLO can be produced by gas reservoirs well below the nominal value of $M_\mathrm{gas}\sim{}50\,\msun$ from \S \ref{sec:continuum}.

\section{Where is it?}
\label{sec:where}

While the MUBLO is seen in projection close to the Galactic Center, only 5\arcmin~from Sgr A*, its line-of-sight location has to be determined.
In this section, we cover the evidence that it is genuinely in the Galactic Center, likely 10-100 pc from Sgr A*.

\subsection{Line-of-Sight Location}
\label{sec:los}
The line-of-sight velocity $v_\mathrm{LSR} \approx 40-50$~\kms is similar to other Galactic Center objects.
There are absorption lines seen in front of the broad lines, and these are clearly from Galactic Center gas based on their continuity with clouds that are definitely in the CMZ (the ``three little pigs'' and the 50 \kms cloud).
Assuming these are genuine absorption features and not interferometric artifacts (it remains difficult to be entirely certain that all interferometric artifacts have been removed, even when working with combined 12m + 7m + TP data), the source cannot be in the foreground of the CMZ.
These absorption features are discussed further in Appendix \ref{appendix:totalpowerspectra}, which shows the ACES single-dish ``Total Power'' data extracted from the same position overlaid on the spectrum we have already shown from the 12m data.
While it is clear that the MUBLO is behind these Galactic Center clouds, it is possible that it could be in the far side expanding 3 kpc arm \citep{Dame2008}, which exists at a similar velocity.  

\begin{figure}
    \centering
    \includegraphics[width=0.48\textwidth]{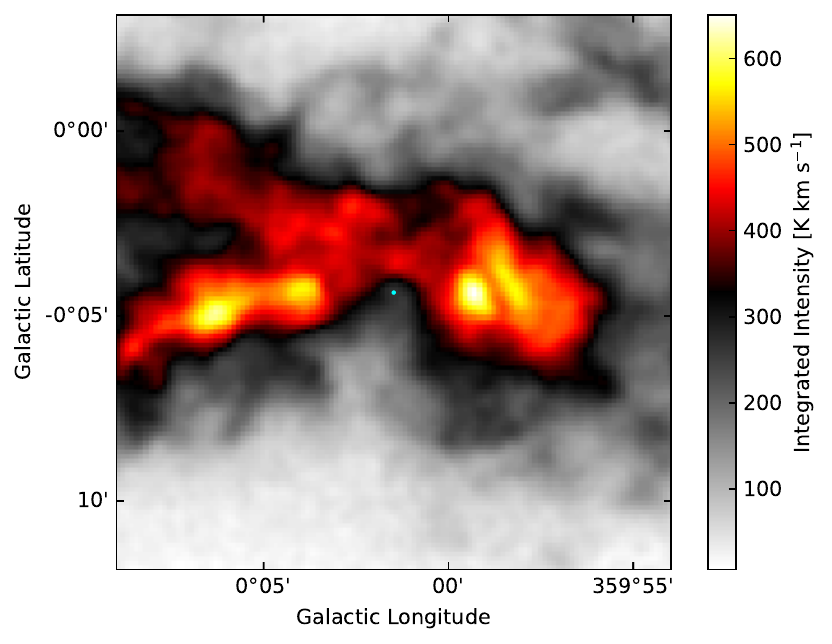}
    \includegraphics[width=0.48\textwidth]{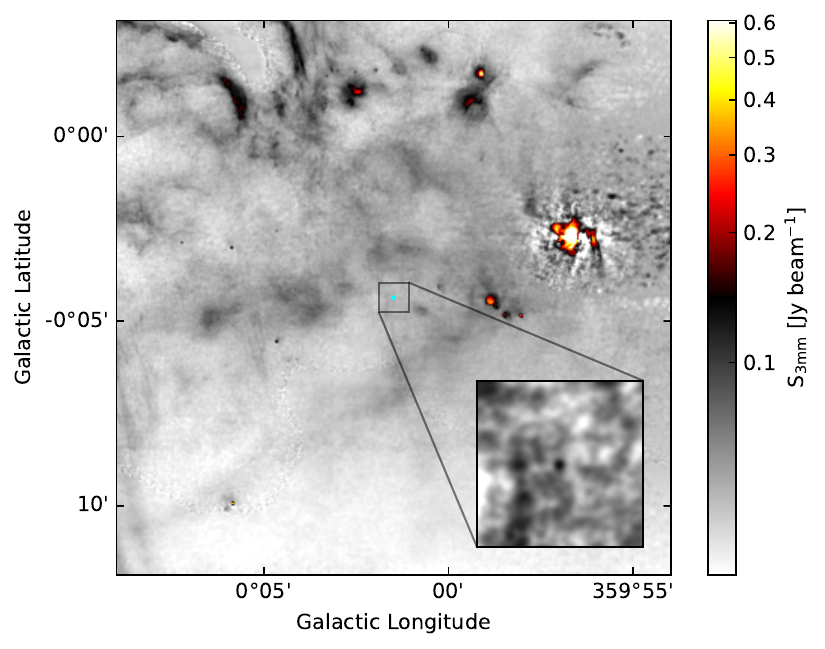}
    \caption{Large-scale images to provide context for the source.
    The left image shows $^{12}$CO integrated intensity from 45 to 55 \kms from the Nobeyama Galactic Center survey \citep{Tokuyama2019}.
    The MUBLO \rr{3 mm continuum} is shown in cyan contours \rr{at 40 and 80 mJy beam$^{-1}$} at the very center of the image.
    The clouds to the left are the `three little pigs' \citep{Battersby2020}, and to the right is the 50 \kms cloud \citep{Uehara2019}.
    The right image shows the MUSTANG combined with ACES \threemm \rr{data}.
    \rr{The inset shows the same data as the parent image with higher contrast to emphasize the MUBLO.}
    The bright source toward the right is the minispiral, which contains Sgr A*.
    The Arched Filaments can be seen in the upper left.
    }
    \label{fig:bigcontext}
\end{figure}

\subsection{Spatial Location}
\label{sec:spatiallocation}
\begin{figure}
    \centering
    \includegraphics[width=0.49\textwidth]{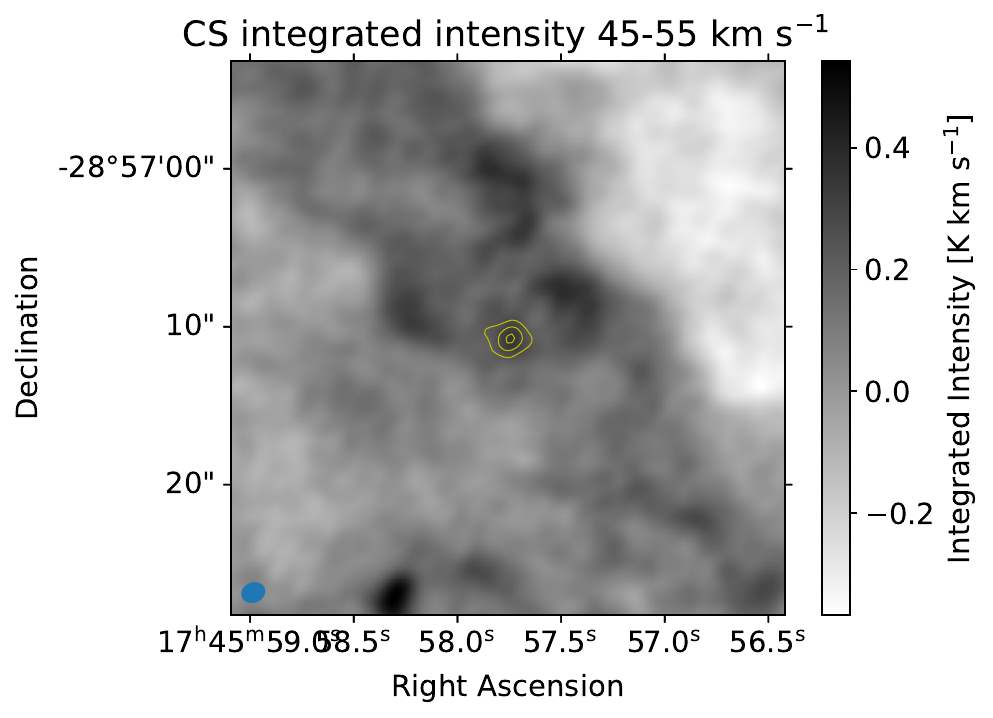}
    \includegraphics[width=0.49\textwidth]{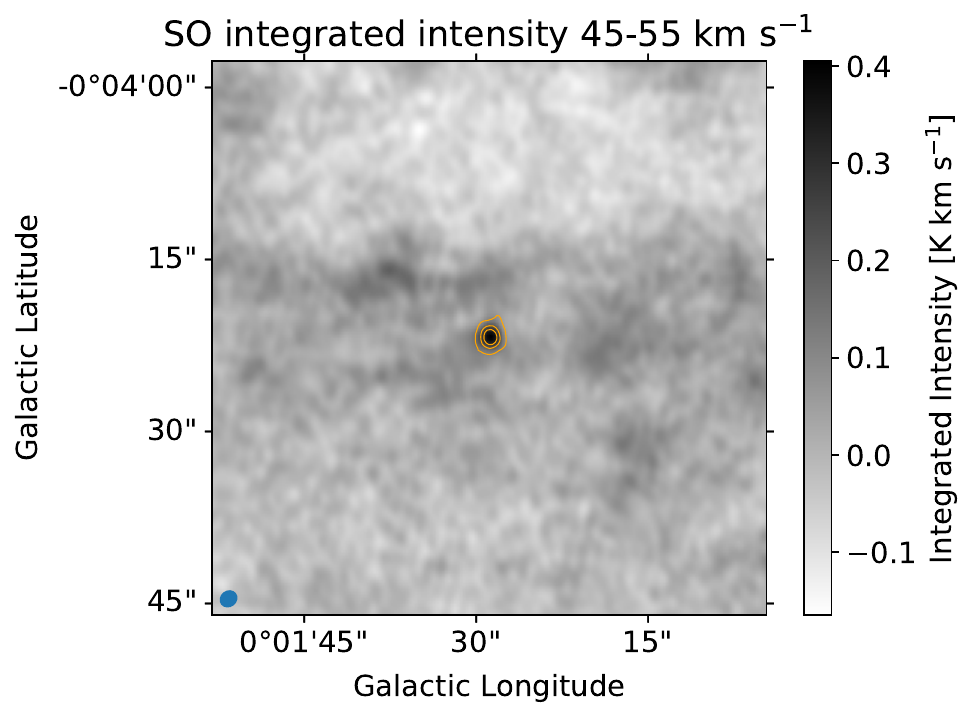}
    \caption{ In both images, the grayscale shows the integrated intensity over the range 45-55 km/s.
    The left image shows CS 2-1 and the right shows SO 2(3)-1(2).
    The contours show the MUBLO integrated over the full velocity range, but with contaminated velocities masked out; the contours are at 20, 40, and 60 K \kms (CS; left) and  20, 60, and 100 K \kms (SO; right).  
    These images provide context on where the compact source resides.
    The MUBLO is not detected in CS over the narrow velocity range because the ``diffuse'' molecular gas at that velocity dominates over the compact source, while the MUBLO is still well-detected in SO over the same velocity range, presumably because the SO 1(2) level (the lower state of the SO 2(3)-1(2) transition) is un- or under-populated in the diffuse cloud.
    }
    \label{fig:sourceoncloud}
\end{figure}

Figure \ref{fig:bigcontext} shows where the MUBLO resides in the large-scale context of the CMZ, showing both CO and \threemm continuum images.
On these larger scales, the MUBLO resides in an underdensity or cavity in the CO gas \citep{Tokuyama2019}.
Figure \ref{fig:sourceoncloud} shows the object in its local context from the ACES data, indicating that there is surrounding molecular gas but that this gas is not particularly associated with the MUBLO.
The gas seen in Figure \ref{fig:sourceoncloud} is sparser than in the neighboring dense clouds seen in Figure \ref{fig:bigcontext}; it is the wispy edge of the clouds seen on the larger scales.

\begin{figure}
    \centering
    \includegraphics[width=\textwidth]{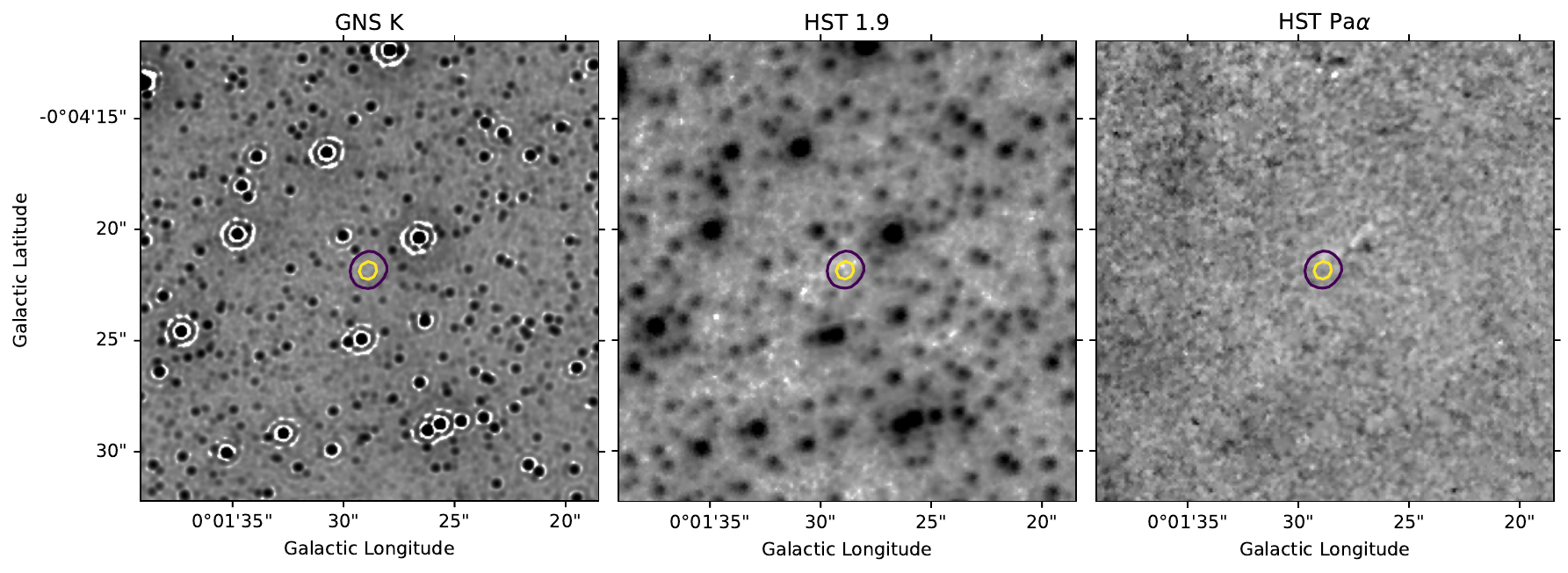}
    \includegraphics[width=\textwidth]{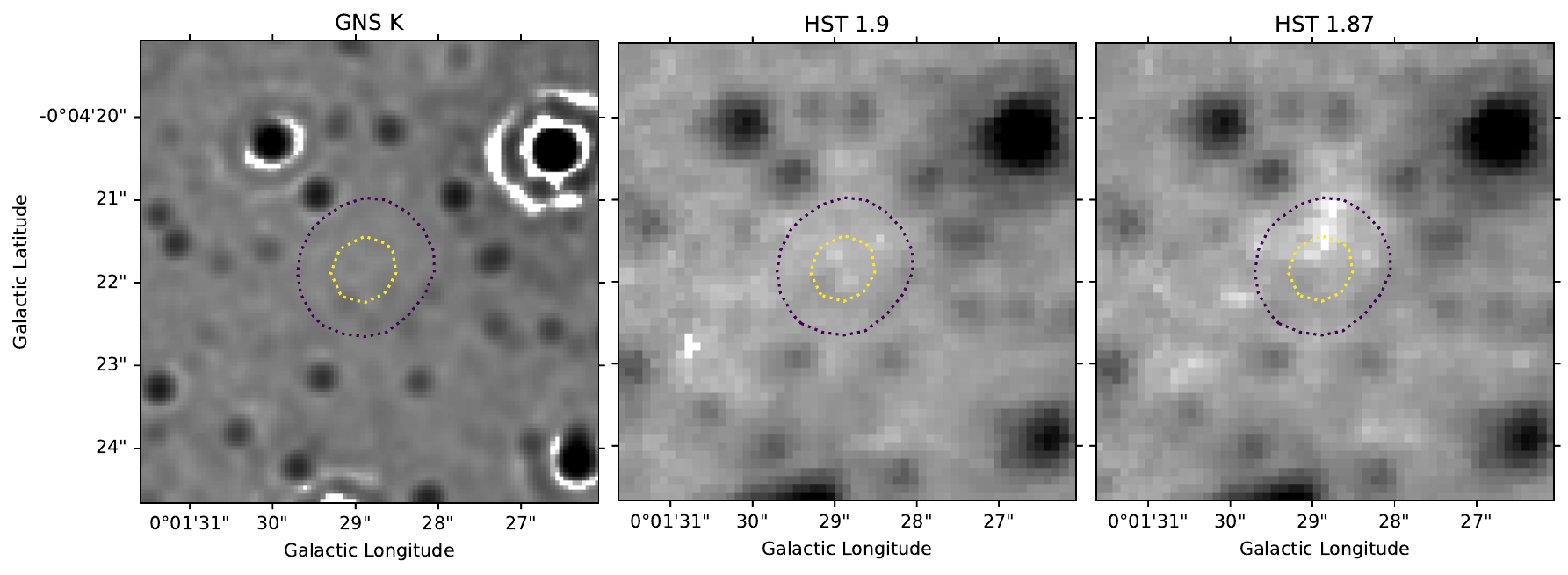}
    \caption{Near-infrared continuum images with mm continuum contours from the ACES 12m data overlaid at 40 and 80 mK.
    (top row)
    Left is the K-band image from GALACTICNUCLEUS \citep{Nogueras-Lara2018,Nogueras-Lara2019}, middle is the F190 filter from the Hubble Space Telescope \rr{(HST)},
    and right is the F187-F190 Paschen $\alpha$ from HST \citep{Dong2011}.
    All are shown with darker color indicating brighter emission.
    There is a hint of Pa$\alpha$ absorption on the north end of the source and just to the northwest of the source, but it is unclear how much to trust these; the bright star to the northwest may produce the apparent absorption as an artifact of the continuum subtraction.
    (bottom row) The left and middle panels are the same, just zoomed further in.
    The right panel is the HST F187 filter, which is the narrow-band that contains the Pa$\alpha$ line. 
    }
    \label{fig:nircontinuum}
\end{figure}

We next check multiwavelength archival data for any counterparts to this source.
Figure \ref{fig:nircontinuum} shows contours from the ACES data overlaid on high-resolution NIR images from the GALACTICNUCLEUS \citep{Nogueras-Lara2018,Nogueras-Lara2019} and HST \citep{Dong2011} surveys.
No source is evident in the NIR data at the location of the MUBLO.
If anything, there is a hint of a deficit of flux at the position of the MUBLO in the HST images, which could be caused by the dust extinguishing background sources.
Archival images from the Hubble Legacy Archive in the F127M, F139M, and F153M filters show this feature more distinctly \citep{Whitmore2016}.
We also checked the surrounding sources from \citet{Shahzamanian2022} within $r<10\arcsec$ and saw no obvious pattern in the proper motions of nearby sources that might indicate a recent runaway or a particularly deep potential well.
We searched Vizier \rr{\citep{Ochsenbein2000}} for any public catalogs with a source at this location at any wavelength and found no compelling counterparts.
The closest sources listed in any catalog are $>1$\arcsec~ away, and these can readily be seen to be outside the contours of the MUBLO in Figure \ref{fig:nircontinuum}.

\begin{figure}
    \centering
    \includegraphics[width=\textwidth]{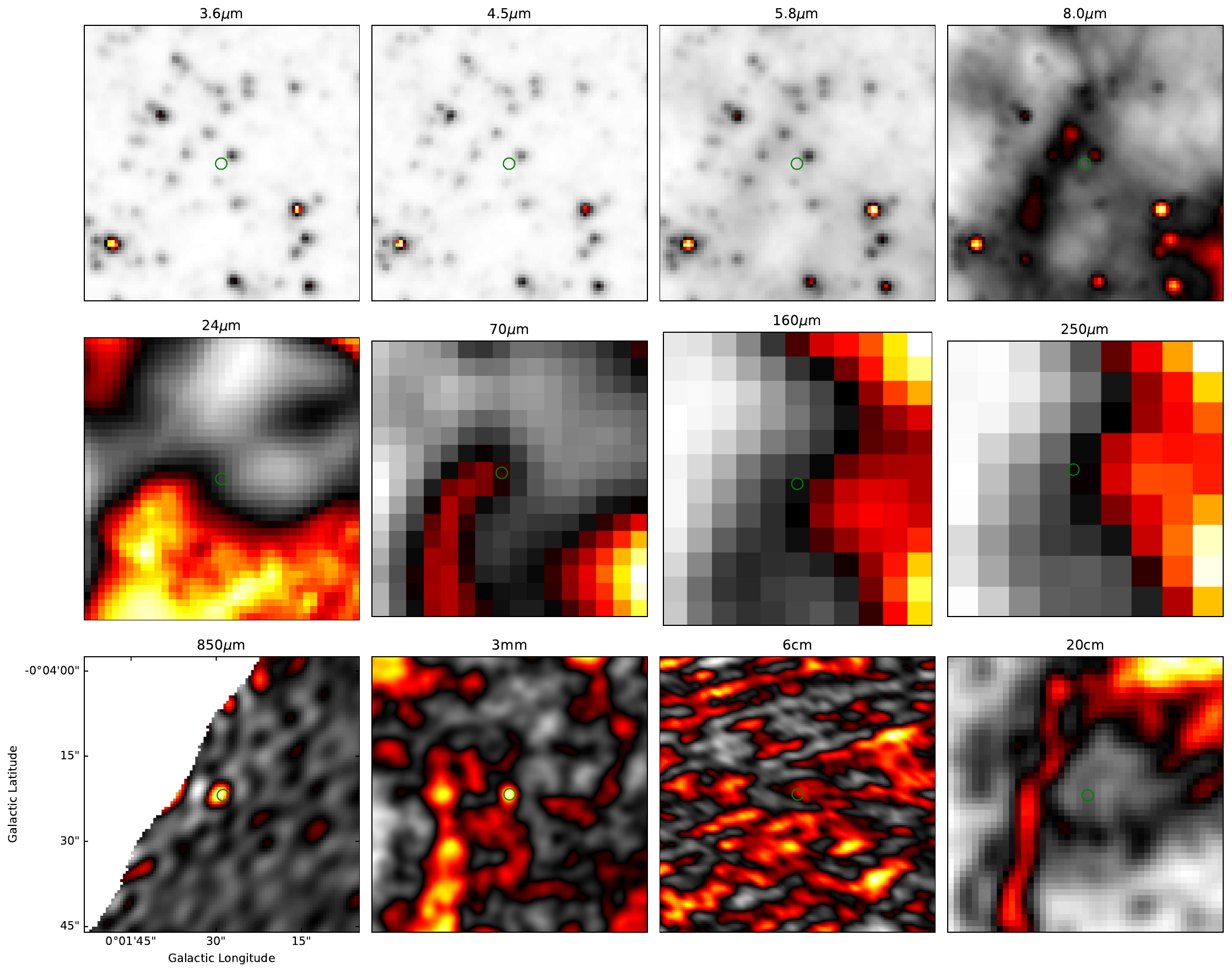}
    \caption{Continuum images of the cutout region at several wavelengths as labeled.
    The cutout is centered on the source coordinates ICRS coordinates 17:45:57.75 -28:57:10.77 and is 35\arcsec on a side.
    The source location is indicated with a green circle, and it is only detected in the two ALMA bands.
    Top row: GLIMPSE \citep{Churchwell2009}.
    Middle row: MIPS \citep{Carey2009} and Herschel \citep{Traficante2011}.
    Bottom row: ALMA (this work), VLA \citep{Lu2019}, and MEERKAT \citep{Heywood2022}.
    }
    \label{fig:contcutouts}
\end{figure}

We further checked longer-wavelength data.
Figure \ref{fig:contcutouts} shows this source in context, with cutouts of the GLIMPSE \citep{Churchwell2009}, MIPSGAL \citep{Carey2009}, HiGal \citep{Molinari2010}, VLA C-band \citep{Lu2019}, and MEERKAT \citep{Heywood2022} surveys.
The only detections are with ALMA at \threemm and 0.85 mm.
This source is unfortunately not covered by the mm-wavelength survey CMZoom \citep{Battersby2020}.

No X-ray sources are present at this location.
The closest cataloged source is $>7\arcsec$ away \citep{Muno2009}.
We have examined recently taken Chandra X-ray data (Vikhlinin et al.~in prep.) comprising 708 ksec effective integration at this location.
The 3$\sigma$ upper limit is $<2.8\times10^{-5}$ counts/s in the 3-8 keV band, equivalent to $S_{X}<7\times10^{-16}$ erg \pers \persc assuming a $\gamma=2$ power-law spectrum.

\clearpage
\section{What is it?}
\label{sec:WhatIsIt}

We have demonstrated the existence of a dusty, broad-linewidth source that is detected only at millimeter wavelengths.
Given this limited information, we now attempt to classify the object.

We consider many options.  Plausible mechanisms include:
    protostellar outflow,
    explosive outflow,
    protostellar inflow,
    ejecta from an evolved star,
    (pre)-planetary nebula,
    stellar collision,
    high-velocity compact cloud (HVCC),
    intermediate-mass black hole (IMBH),
    galaxy,
    or supernova.
We evaluate each of these hypotheses in the following sections, but find that none satisfactorily explain the data.

\subsection{Something associated with star formation}
\label{sec:starformation}
Star formation is prevalent in the Galactic Center, with roughly 10\% of the Galaxy's star formation occurring in the central $r<100$~pc \citep{Longmore2013,Barnes2017,Henshaw2023}.
We therefore evaluate several hypotheses associated with star formation.
Star-forming regions are naturally dust- and molecule-rich.
We consider whether the MUBLO is a protostar (\S \ref{sec:protostar}), a typical protostellar outflow (\S \ref{sec:protostellaroutflow}), an explosive outflow (\S \ref{sec:explosion}), a protostellar inflow (\S \ref{sec:inflow}), a protostar collision (\S \ref{sec:protostarcollision}), or a prestellar core collision (\S \ref{sec:corecollision}).
While these hypotheses can explain some of the bulk properties of the MUBLO, they all fail in the details, particularly energetics, morphology, and chemistry.

\subsubsection{Protostar}
\label{sec:protostar}
Before digging into the specific models to explain the gas, we evaluate what kind of central source is allowed by the observed dust SED described in \S \ref{sec:continuum}.
Using the \citet{Richardson2024} update to the \citet{Robitaille2017} grid, we searched for models consistent with either of the two ALMA millimeter detections (we did not require that both measurements match the models since the model grid does not allow for a varying $\beta$ parameter).
We search for \rr{models} that have flux density within $5000$~au apertures \rr{matching the ALMA measurements to within 25\% and falling below the 24 and 70 \um upper limits}.
There is a narrow range of parameter space compatible with the ALMA measurements and Herschel and Spitzer upper limits, all with $10^4 < L < 10^{4.7}~\lsun$ and $3<M_{\mathrm{gas}}(<5000~\mathrm{au})<28$~\msun.
While these objects are part of the model grid, they are not compatible with most models of star formation, since the implied central source is an $M\gtrsim10\msun$ star that is surrounded by a comparable amount of cold gas in a stable envelope-plus-disk configuration.
Instead, these models demonstrate that there exist solutions in which high-luminosity, but still $L<10^5$ \lsun, stars may be embedded in dusty envelopes that produce enough millimeter-wavelength emission while not exceeding the short-wavelength limits.

\subsubsection{Protostellar outflow}
\label{sec:protostellaroutflow}
The hypothesis that the object is a protostellar outflow from a previously-unknown star-forming region is plausible.
However, there is a good deal of evidence that suggests this is not the correct interpretation\rr{:}
\begin{enumerate}
    \item  The limited \rr{number of} emission lines detected, and especially the lack of SiO, would make this outflow unlike any others in the Galaxy.  Admittedly, our sensitivity to CO is limited, since it was only covered in the coarse \rr{spatial} resolution B7 data \rr{such that diffuse molecular gas along the line of sight is confused with and may absorb any compact emission associated with the MUBLO}.
    \item It is surprisingly compact, $<10^4$ au, which suggests that it is almost perfectly face-on (any angle would produce an extended, red/blue bipolar signature).  Similarly, the \rr{small value of the spatial gradient with velocity} suggests that it is face-on.  This situation is unlikely, but possible.  
    \item The line profile is Gaussian.  Such a profile is unexpected for a straight-on outflow, which would more likely include some sharper features or flattening from the high velocity tails.  However, if the material we observe is entirely entrained ISM, a Gaussian profile is not impossible.
    \item Assuming this is a face-on outflow-driving source, we would expect to see the central YSO \rr{(young stellar object)} at infrared wavelengths.  The nondetection at wavelengths short of 850 $\mu$m suggests either that there is a very optically thick dust envelope or that the driving source is low-mass and low-luminosity.
    \item The source appears isolated, with no surrounding sources at any wavelength, and very little surrounding cloud material.  YSOs are not often found in such environments.
    \item High-velocity outflows are expected to produce high-velocity shocks in the dense interstellar medium gas.  The lack of SiO emission and the low inferred temperature both imply that there are no high-velocity shocks.  This evidence is the most problematic for the protostellar outflow hypothesis. 
    \item Such massive (50 \msun), compact ($<10^4$ au) cores are rarely observed and are generally much hotter \citep[e.g.,][]{Bonfand2017,Bonfand2024,Jeff2024,Budaiev2024}.
\end{enumerate}

A protostellar source is not strictly ruled out, but is very unlikely.

\subsubsection{Explosive outflow}
\label{sec:explosion}
Could this be an explosive outflow at an early stage analogous to Orion BN/KL \citep[e.g.,][]{Bally2015,Bally2017}?

For this hypothesis to hold, the explosion must be very young, with $t\approx 10^4$ au / 70\kms$\approx$700 yr.
An isotropic explosion is consistent with the line profile.

This hypothesis has several problems in common with the protostellar outflow hypothesis: there is little surrounding interstellar medium, there is no associated infrared emission (especially near-infrared H$_2$ and [Fe II], which would show up in K-band), there is no shocked SiO emission, and the excitation \rr{temperature is very low}.
The Orion BN/KL outflow is bright in all of the\rr{se} features that are not detected toward the MUBLO.

\subsubsection{Protostellar Inflow}
\label{sec:inflow}
Could this be a site where large amounts of material are inflowing toward a central, collapsing source?
This hypothesis explains the lack of infrared detection, but otherwise fails to explain all the main observables.
In particular, the huge linewidth is unexpected for a collapsing source unless it is extremely massive (see \S \ref{sec:imbh}).
We would also expect to see an accretion shock, which might be expected to produce hot line emission.

\subsubsection{Protostellar Collision}
\label{sec:protostarcollision}

Protostars are likelier to collide than their main-sequence counterparts.
They are bloated during their early phases as they radiate away their gravitational energy, and they reside in a dissipative medium that can result in multiple systems inspiraling.
A protostellar collision would appear similar to a later-stage stellar merger as described in \S \ref{sec:merger}.
We defer further discussion to that section.

\subsubsection{Prestellar Core Collision}
\label{sec:corecollision}

Gas falls in along the Galactic bar at high velocity, impacting the central molecular zone at high speeds \citep{Sormani2019,Gramze2023}.
If dense prestellar cores were to form along the Galactic bar's dust lanes and impact one another in the center of the Galaxy, the high observed velocity dispersion could be produced.
However, in doing so, extremely strong shocks should occur, and therefore we would expect to see bright SiO emission.
Furthermore, this scenario is intrinsically unlikely, as the free-fall timescale for a prestellar core with half the mass of the MUBLO (assuming two equal-mass cores, the most conservative limit) is only 25 kyr, so these cores would have had to have formed extremely recently.

\subsection{Evolved Star}
\label{sec:evolvedstars}
Could this be some sort of evolved star, such as an asymptotic giant branch or red supergiant star with an extreme wind?
These mass-losing stars are generally detected in emission from sulfur-bearing species \citep[e.g.,][]{Omont1993}.
The main evidence against this hypothesis is the lack of an infrared source, though the lack of an SiO $v$=1 maser \rr{in the ACES data} is also weak evidence against the MUBLO being an evolved star \rr{\citep[at least 15\% of red supergiants exhibit SiO masers;][]{Verheyen2012})}.
These end-of-life stars are generally extremely luminous; the known red supergiants in the Galactic center have observed K-band magnitudes $m_K < 6$ \citep{Schultheis2020}.

It is theoretically possible that one could be hidden by a very high column density of dust produced in its own wind, similar to the R Coronae Borealis stars and OH/IR red supergiants like VY CMa \citep{Humphreys2024}, or episodic mass loss leading to events like Betelgeuse's Great Dimming \citep{Montarges2021,Levesque2020}, but the high column density required for this mechanism to completely block the star's infrared light would make the MUBLO unique.
The required local extinction \rr{must} be $A_K>10$ \rr{for an $m_K=6$ star} to be undetected in the GALACTICNUCLEUS data \citep{Nogueras-Lara2019}; while this amount of dust is compatible with the observed millimeter-derived column density (\S \ref{sec:continuum}), a $10^5$ \lsun star would heat the dust well above the upper limit of $T_D<50$ K (the protostar models discussed in \S \ref{sec:starformation} demonstrate this point).

VY CMa is a helpful reference case as perhaps the most extreme mass-losing supergiant star in the Galaxy.
It is a $L\sim2\times10^5$~\lsun \citep{Monnier1999} star with $M>2.5\times10^{-4}$~\msun of circumstellar dust \citep[$M>2.5\times10^{-2}$~\msun of gas using our gas-to-dust ratio;][but see \citealt{Kaminski2019}, who model dust mass as much as 100$\times$ higher]{O'Gorman2015}.
If we were to consider VY CMa as only a millimeter continuum source, it is similar to the MUBLO: its 350 GHz flux, scaled to a distance of 8 kpc, is 40 mJy \citep{O'Gorman2015,Kaminski2013}, within a factor of two of the MUBLO.
Similarly, its 100 GHz flux (which we \rr{extrapolate} from Figure B.1 of \citealt{O'Gorman2015}) scaled is 2 mJy, about what we measure.
However, the dust temperatures they measure are $>10\times$ hotter than allowed by our observational limits.
VY CMa is a bright IRAS source, with $S_{25\um}=149$ Jy \citep[scaled to $d$=8~kpc,][]{1988iras....7.....H,Matsuura2014}, which is $>100\times$ the Spitzer MIPS upper limit.

Additionally, the stars that produce the most dust in their winds tend to drive slower winds, while hotter stars that drive faster winds, compatible with the $>70$ \kms we observe, tend to have less massive winds \citep[e.g., the fastest winds in a sample of mass-losing giants, VY CMa and IRC+10240, have FWHM$\sim60$~\kms, less than half of the MUBLO's;][]{Kemper2003,Quintana-Lacaci2023}.
Most of the mass is in lower-velocity material, with $v_{max} < 40~ \kms$ \citep{Quintana-Lacaci2023}.
The molecular winds in these sources are bright in SO, like the MUBLO, but comparably bright in SiO, which the MUBLO is not \citep{Kaminski2013,Matsuura2014,Quintana-Lacaci2023}.

The dust mass we infer requires a truly extreme star to reproduce.
The only star we know of that has ejected $\sim10\msun$ of matter outside of a supernova is $\eta$ Carinae, whose great eruption produced $\sim10~\msun$ in only a few years.
However, $\eta$ Car bears no observational similarities to the MUBLO.
It has narrow molecular lines \citep{Bordiu2022} because most of the ejecta are ionized.
The system is also far too bright; its distance-scaled flux is $S_{\mathrm{100 GHz}}\approx3$~Jy \citep{Morris2020}, over $10^3\times$ greater than the MUBLO.

\subsubsection{(Pre-)Planetary Nebula}
Following along the evolved star route, could this be a star that has evolved past the point of nuclear burning, traveling into the pre- or planetary nebula phase?
The planetary nebula hypothesis is unlikely, since there is no sign of ionized gas, but pre-planetary nebulae can be much higher density and cooler.
The limited \rr{number of} molecules detected could be a consequence of some peculiar enrichment process in the star, though we have no model for such a process.

The line width is one possible problem with this hypothesis.
In observed PNe/PPNe, the core line width in molecular gas is generally small \citep[e.g., CRL 618 has widths $\sim15$ \kms;][]{Lee2013a,Lee2013b}.
There are examples of broad lines, though: the Boomerang nebula has $\sim100$ \kms wide absorption from its fastest-expanding material seen in CO \citep{Sahai2017}.
There are also many $\sim$150 \kms molecular jets detected among (P)PNe \citep{Guerrero2020,Sahai2015}.
These jets tend to be much fainter broad-wing components next to a central, more massive component, and in at least some cases they have coincident SiO, so \rr{the MUBLO} would still stand out as unique, but there are analogues.

In one of the most extreme molecular PPN examples, I08005,  which has 200 \kms wide CO and SiO lines, no continuum was detected at an upper limit of 1 mJy at 870$\mu$m at a distance of 3 kpc \citep{Sahai2015}.
The detection of the MUBLO at 90 mJy at 8 kpc at the same wavelength makes it $>400\times$ brighter in the mm continuum than this PPN.
This difference suggests that the MUBLO is too dust-rich to be a PPN.

The lack of a detected continuum source at short wavelengths is again a problem for the PPN hypothesis.
For a PN, we would not necessarily expect a central continuum source to be detectable in the infrared, but PPN usually have fairly large stellar photospheres and are luminous in the infrared.

While there are some similarities between (P)PN and the MUBLO, most of the evidence suggests that these are not the same class of object.

\subsubsection{Stellar Merger: A Luminous Red Nova}
\label{sec:merger}

Could the object be the result of stellar merger?
Luminous Red Novae (LRN) are a class of transients thought to be associated with stellar mergers, and the MUBLO shares some observational features with the \rr{remnants of these events}.
Stellar mergers are expected to be more common in the high-density inner Galaxy, and may even be the origin of the G objects near the Galactic Center \citep{Ciurlo2020}.
Stellar mergers that produce luminous red novae are often accompanied by high-velocity, cold molecular outflows \citep{Kaminski2018}.
The energy released in stellar merger events can be $\sim10^{48}$ erg \citep{Retter2006}, comparable to the energy in the MUBLO (\S \ref{sec:continuum}), though the component in the molecular remnant of these mergers is $\sim10^{46}$ erg \citep[][]{Kaminski2018}.
The very large energy in the molecular gas in the MUBLO, $E\sim5\times10^{48}$ erg, suggests that a complete merger rather than a glancing collision is more likely; the gravitational energy in a merger of two solar-mass objects is $4\times10^{48}$ erg.

Among the handful of known Galactic LRN, four with mm/submm spectral line observations exhibit SO/SiO ratios ranging from 1 to 7, as reported in studies by \citet{Kaminski2015,Kaminski2018,Kaminski2020}.
These ratios are notably two orders of magnitude lower than the ratio observed towards MUBLO.
\rr{Gas temperature} estimates from \rr{SiO, SO, and SO$_2$} for three of these four sources exceed $T>50$ K, with \citet{Kaminski2018} identifying temperatures above 200 K in SO$_2$, which is substantially higher than the temperature estimates for MUBLO.
CK Vul, the oldest RN, stands out as the only exception, with a temperature of 12~K.
Therefore, the chemistry and excitation conditions of MUBLO are  different from the handful of known RNs.

We explore two examples, V838 Mon and CK Vul, in more detail.

\paragraph{V838 Monocerotis}
V838 Mon exhibits some similarities to the MUBLO.
ALMA observations of V838 Mon 17 years after outburst reveal broad line widths, $v_{\mathrm{fwhm}} \approx 150~\kms$, in lines of CO, SO, SiO, and AlOH spread over $\sim 700$ au \citep{Kaminski2021}.
The integrated intensity of the SO 5(6)-4(5) line in their data is $S\sim9$ Jy \kms, which, if converted to a distance of 8 kpc, would drop to $\sim4$ Jy \kms ($d_{V838}=5.9$ kpc), comparable to the observed integrated intensity of SO 2(3)-1(2) in the MUBLO (Table \ref{tab:spatial_measurements}).
However, they observe an SiO 5-4 / SO 5(6)-4(5) ratio $>2$, while our object has SiO 2-1 / SO 2(3)-1(2) $<0.02$, implying a dramatically different chemistry is present.
The continuum observed toward V838 Mon at 1mm is $S\approx2$ mJy (1 mJy at 8 kpc); for comparison, interpolating our observed flux between the B3 and B7 observations, the expected flux of the MUBLO is $S_{1mm}=30$ mJy, so about 30 times brighter.

V838 Mon is also a notably bright infrared source.
From 2010 to 2020, it remained at $m_K \lesssim 5$ \citep{Woodward2021}, which would be roughly $m_K<6$ at \rr{d=8 kpc}.  
The K-band upper limit from GALACTICNUCLEUS is about $m_K > 16$ \citep[the 80\% completeness limit;][]{Nogueras-Lara2020}, so $A_K>10$ would be required to hide \rr{a central source like} V838 Mon in the MUBLO (and \rr{extinction along the Galactic plane is only} $A_K\sim3$ \rr{toward} the Galactic Center).
In the mid-infrared, V838 Mon is bright, $S_{19.7 \um}=38$ Jy (18 Jy at $d_{\mathrm{GC}}$), while the upper limit from Spitzer 24 \um~toward the MUBLO is about 1.3~Jy, which requires $A_{24 \um}\approx3$ to hide.
At longer wavelengths, V838 Mon is fainter and the limits from Herschel are less stringent.

\paragraph{CK Vulpecula}
A second comparison source of interest is CK Vulpecula, a \rr{LRN} that occurred in 1670.
Unlike V838 Mon, its SED is a reasonable match to that of the MUBLO.
It has long-wavelength fluxes that are comparable to those of the MUBLO; assuming $d_{\mathrm{CK}}=3.2$ kpc \citep{Banerjee2020,Kaminski2021_CKII}, its fluxes scaled to $d=8$ kpc are $S_{\mathrm{100 GHz}}\approx2$ mJy, $S_{350 GHz}\approx20$ mJy, $S_{24 \um}=1.5$ mJy, and $S_{6 GHz}=0.2$ mJy \citep{Kaminski2015}.
The millimeter measurements are within a factor of a few of the MUBLO's measurements, and the others are consistent with the upper limits: CK Vul does not have a detected central source at infrared ($<24$ \um) wavelengths.
The preferred SED model in \citet{Kaminski2015} has dust with $T=15$~K and $\beta=1$, with inferred central source luminosity\footnote{\citet{Kaminski2015} reported $L=1\lsun$, but that work adopted $d_{CK}=0.7$~kpc; later works have all confirmed that it is more distant \citep{Banerjee2020,Kaminski2021_CKII}} $L\approx20$~\lsun, and these authors infer a total gas mass of $M=1 \msun$  from their CO observations (they do not report a dust mass).

The key differences between CK Vul and the MUBLO are in its emission lines.
All of the molecules seen in the MUBLO, SO, SO$_2$, and CS, are detected in CK Vul \citep{Kaminski2020_CKI}.
However, the SO and SO$_2$ lines in CK Vul are $\gtrsim10\times$ fainter than the SiO lines that are not detected in the MUBLO, and several lines of SO seen in the MUBLO are not detected in CK Vul because they are swamped by transitions from other molecules, e.g., HC$^{15}$N, that are not present in the MUBLO \citep{Kaminski2017}.
While there is significant emission in CK Vul with FWHM$\sim100-200~\kms$ seen in CO and CS, the SO$_2$ lines are narrow, $\lesssim50~\kms$ \citep[][SO is detected but its line profile is not shown]{Kaminski2020_CKI}.
The abundances of SO, SO$_2$, CS, SiO, and HC$_3$N are all roughly the same (equal within errorbars) in CK Vul \citep{Kaminski2020_CKI}, in contrast to the MUBLO, where \rr{SO and SO$_2$ are much more abundant than SiO and HC$_3$N}.
In summary, while CK Vul is quite similar to the MUBLO in the continuum, it is dramatically different in molecular lines.

\vspace{0.5em}

The stellar merger / luminous red nova hypothesis seems quite plausible, but there remain several features that distinguish the MUBLO from other LRN.
The dust mass is substantially larger, by more than an order of magnitude, than observed toward any other merger remnant.
There is no hint of a central source at infrared wavelengths, making the MUBLO much more obscured than any previous LRN except CK Vul.
The chemistry is dramatically different, with the MUBLO characterized by a lack of SiO.
Together, these arguments imply that if the MUBLO is a merger remnant, it was from a merger of more massive stars than previously-observed LRN (to account for the extra mass), and it occurred $>10$ years ago to account for the re-freezeout of SiO \rr{(but in old LRN, like CK Vul, SiO has not frozen out)}.
The nova itself should have been extremely luminous, then, and have driven light echoes comparable to those created by V838 Mon - these might then be detectable in scattered light in the infrared if the event was recent enough.

\subsection{High Velocity Compact Cloud}

There have been many high-velocity compact clouds (HVCCs) reported in the Galactic Center region.  These are peculiar clouds characterized by their compact sizes ($d\lesssim 5$ pc) and broad velocity widths ($\Delta V\gtrsim 50$ \kms).  These have been explained as either material accelerated by supernova explosions, a connecting bridge between colliding clouds, or gas orbiting around invisible massive objects \citep{Oka2014,Oka2016,Oka2022,Iwata2023}.  The MUBLO shares some properties with HVCCs, specifically the broad line width, but it is much smaller than the known HVCCs.  All reported HVCCs have been found with single-dish telescopes, and thus are extended over parsec scales, while the MUBLO has \rr{a radius smaller than} $r<5000$ au ($r<0.02$ pc).  Recent ALMA observations toward two HVCCs  detected several unresolved ultracompact clumps (UCCs) with broad velocity width  \citep[$\Delta V\sim 50$ \kms;][]{Takekawa2019,Iwata2023}.
These differ from the MUBLO in a few observational respects: their line widths are somewhat narrower, they are detected in different lines (CO, CH$_3$OH, SiO, HCN), they do not contain compact millimeter continuum sources, and they are surrounded by and connected to extended high-velocity-dispersion gas.
They are therefore unlikely to be the same class of object.

\subsection{Intermediate-Mass Black Hole (IMBH)}
\label{sec:imbh}
Given the broad linewidth, it is possible that the MUBLO is comprised of gas in orbit around a very deep potential well.
Because of the nondetection at multiple wavelengths, that potential well is dark, so a cluster of stars is an unlikely explanation.
We therefore consider whether the object may be an intermediate-mass (of order $10^4$ \msun) black hole.

There have been many previous claims of IMBH detections in the CMZ \citep{Oka2016,Tsuboi2017,Takekawa2019,Takekawa2020}.
These have been hard to confirm, since there exist many alternative explanations for broad-linewidth gas \citep{Ballone2018,Oka2017,Ravi2018,Tanaka2018}.
We therefore approach this hypothesis cautiously, recognizing that significant evidence is required to claim that a black hole is the only acceptable explanation.

\begin{figure}
    \centering
    \includegraphics[width=0.5\textwidth]{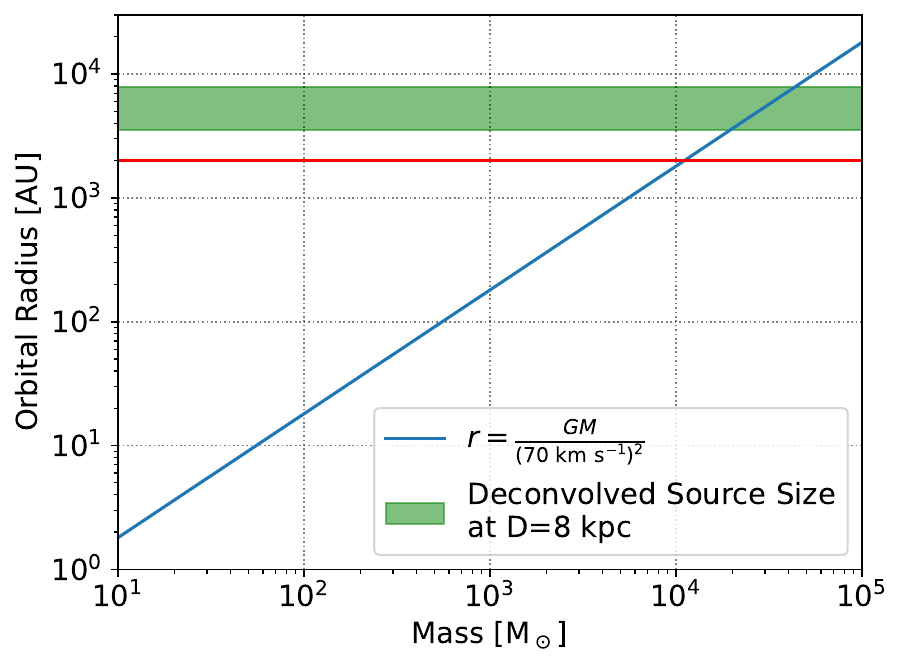}
    \caption{For the IMBH model, we plot $r=GM/v^2$ to obtain limits on the mass assuming the source is resolved (see Table \ref{tab:spatial_measurements}).  The green highlighted zone covers the range from the minor to major deconvolved source size.
    The red line at 2000 au shows the location of the fitted offset between the red and blue lobes of the SO line (Figure \ref{fig:SOredblue}).
    }
    \label{fig:imbhorbit}
\end{figure}

The velocity profile observed has a width that could be produced by an orbit around a central potential.
An orbital velocity of 70-80 \kms (roughly the half-width of the MUBLO's line profile) would occur at $10^3-10^4$ au for a few $10^4$ \msun black hole (see Figure \ref{fig:imbhorbit}).
However, a disk with a pure Keplerian orbit \rr{seen edge-on} should not produce a Gaussian profile, but should instead produce a double-peaked profile with peaks corresponding to the disk's \rr{outermost} radii.
The peak of the SO profile is a key limit on this model, since it appears consistent with a smooth Gaussian (e.g., Figure \ref{fig:coarse_spectra}).
In order for the double-peak profile to be obscured, the purported disk would either need to be face-on, which would reduce the observed line width, or the gas would need to be very turbulent.
A very high degree of turbulence is plausible, but would likely produce significant shocks, and therefore we would expect to see SiO emission.

Alternatively, a lower mass object would produce this broad linewidth at smaller radii.
The apparent low optical depth of the SO and CS lines sets a lower limit on the radius: if we assume the lines are optically thin, the filling factor must be $ff>0.1$ (\S \ref{sec:spectralmeasurements}), limiting the radius to $r>1400$ au, which implies a lower mass limit $M_{BH}>8000$ \msun following Figure \ref{fig:imbhorbit}.
This is not a strict limit, however, as the line profile would not necessarily appear non-Gaussian for moderate optical depth; more detailed modeling would be needed to produce a firm lower limit.

Despite the apparent problems with a disk model, we explore it a bit further.
We assume a central mass $M=10^4$ \msun such that the orbital velocity reaches $\sim70$ \kms at around $2000$ au.
We adopt a simple viscous disk model with $\alpha=0.001$ and total mass 50 \msun, which results in surface density $\Sigma=7\times10^{25}\persc$, and Planck mean opacity $\kappa=3$ cm$^{2}$ g$^{-1}$.
The midplane temperature is then \rr{\citep[][problem set 4]{Krumholz2015}}
\begin{equation}
    T_m = \left[\frac{27  \alpha k_B  \Sigma^2 \kappa \Omega}{64 m \sigma_{SB}}\right]^{1/3}
\end{equation}
which we parameterize as
\begin{equation}
\label{eqn4}
T_m = 7.3 \mathrm{~K~}
        \left[\frac{M}{50~\mathrm{M}_\odot}\right]^{2/3}
        \left[\frac{\kappa}{3~\mathrm{~cm^2~g^{-1}}}\right]^{1/3}
        \left[\frac{\alpha}{10^{-3}}\right]^{1/3}
        \left[\frac{r}{2000~\mathrm{au}}\right]^{-4/3}
\end{equation}
This temperature is within a plausible range to explain the observed $T_\mathrm{LTE}\approx13$ K, though it may imply either higher $\alpha$, a smaller emitting radius, or perhaps some other heating source is present (e.g., cosmic rays).
We also compute the inner radius $r_\mathrm{min}=9$ au \rr{by inverting Equation \ref{eqn4} and} assuming $T_\mathrm{max} < 10^4$ K, since there is no detected ionized gas in the cm continuum or in recombination lines.
The inner radius at which we would expect to detect molecular lines is roughly the dust destruction radius, $T\approx2000$ K, at $r=30$ au.

Additionally, we checked the stellar kinematics in the vicinity of our target source.
\citet{Shahzamanian2022} reported proper motions measured using the GALACTICNUCLEUS data.
Within a few arcseconds of this object, there are several measured proper motions, but there is no \rr{apparent} trend; in particular, there is no increase of PM closer to the source.
This lack is not evidence for or against the IMBH hypothesis, though, as most stars in the field-of-view are likely at large physical distance from the MUBLO despite their small projected distance.

While there are several appealing features of the IMBH model, in its simplest form, it does not explain all of the observed features of the MUBLO.
We regard it as a possible explanation, but do not favor it above other models.

\subsection{Galaxy}
Could this object be a background galaxy?

Assuming we have correctly identified the spectral lines in \S \ref{sec:spectralmeasurements}, the line-of-sight velocity of 40 \kms makes a background galaxy hypothesis very unlikely.
There are few galaxies at this redshift (i.e., near zero) and these galaxies occupy a low density on the sky.
If the MUBLO were a galaxy, we would be observing a very compact component of it, perhaps the center.
Assuming a size scale of the center of $r\sim100$ pc, in order to be unresolved at $1\arcsec$ resolution, it would need to be at $D\gtrsim20$ Mpc.
At that distance, the expected redshift is $H_0 D \sim 1400$ \kms, such that the required peculiar velocity for this to be a galaxy would be $\sim-1300$ \kms, much larger than expected in the local neighborhood.  

The observed chemistry also is evidence against this being a galaxy.
We compare the MUBLO to the CMZ as a whole: taking the average over the whole observations of the GC from ACES Total Power observations, our CMZ has an average CS \rr{2-1} brightness of 53 K \kms and SO \rr{2(3)-1(2)} brightness 10 K \kms.
Table \ref{tab:spectral_measurements} shows that the MUBLO has SO 2(3)-1(2)/CS 2-1 of about 1.8, while the CMZ has a ratio of 0.19, about a 10$\times$ difference.
This other hypothetical CMZ would have to exhibit a dramatically different chemistry to our own.  The HC$_3$N/CS brightness ratio in the MUBLO is $R<0.04$, while the measured value in the CMZ is 0.11 both from our measurements and  \citet{Jones2012}, which is also a discrepancy, but not quite as difficult to reconcile as the SO/CS ratio.

Given that a correct identification of the lines essentially rules out a background galaxy, could we have misidentified the lines?
This possibility is entirely ruled out, as we have identified several lines of SO and its $^{34}$S isotopologue in addition to a CS and C$^{34}$S line.
While any one line could be misidentified, a grouping of five different lines redshifting exactly on top of the rest frequencies of lines of another species is exceedingly unlikely.

\subsection{Supernova}

Could this be the remnant of a star that went supernova, or perhaps even a failed supernova?
A supernova would readily explain the lack of an infrared source, since the purported source would have exploded.

There is at least one recent example of a `failed supernova' candidate in which a source remained afterward, but was highly reddened and fainter \citep{Adams2017,Beasor2023,Kochanek2024}.
As far as we are aware, though, there are no millimeter-wavelength observations of N6946-BH1, so we cannot (yet) make a direct comparison between it and the MUBLO.

The presence of molecules is somewhat compatible with a supernova.
SN1987A exhibits molecular emission in CO and SiO with widths $\sim1000$ \kms \citep{Cigan2019}.
If the MUBLO is a supernova remnant (SNR), it is both much narrower in linewidth and chemically distinct from SN1987A.
Under the SNR hypothesis, we should probably adopt a much lower gas-to-dust ratio such that the total mass of the MUBLO is $\lesssim1$ \msun.
As explained in \S~\ref{sec:weirddust}, recent work suggests the presence of dust mass reservoirs $\sim{}0.5\,\msun$ in some supernova remnants, underlining this point. 

There is weak circumstantial evidence from the morphology of the surrounding material that there was an explosive event at the MUBLO's location.
The gap between the 50 \kms cloud and the Three Little Pigs (Figure \ref{fig:bigcontext}) could be produced by a supernova, in principle.
The MEERKAT and Chandra images of the area give no hint of a supernova remnant, though, so if this was a supernova-driven cavity, its hot gas has already disappeared.
Additionally, the size scale of the apparent gap is several parsecs, while the MUBLO's molecular emission is only a few thousand au, which is difficult to reconcile.

Supernovae are comprised of fast-moving ejecta, and therefore should expand over time.
The line width and the molecular and dust continuum intensity have all stayed roughly constant from 2012 to \rr{2022} (\S \ref{sec:continuum}).
It is not clear how a supernova scenario would explain the steadiness of the observed dispersion over this $\sim10$ year period.
It is possible that we are seeing only the inner remnant that has been confined by the reverse shock.

In summary, we cannot rule out a supernova remnant as an explanation for the MUBLO, but such a model does not account for all of its features.

\section{Conclusion}
\label{sec:conclusion}
We found a source exhibiting extremely high velocity dispersion, which we title a Millimeter Ultra-Broad Line Object (MUBLO).
We do not have a conclusive classification of this source.
Several hypotheses produce many of the observed features, but none explain them all.

The key observational features of the MUBLO are:
\begin{itemize}
    \item It has large linewidth ($FWHM=160~\kms$, $\sigma=70$ \kms) in molecular lines (SO and CS).
    \item It is likely in the Galactic Center, since the broad lines show absorption from Galactic Center clouds.
    \item It is compact, $\theta_{FWHM,maj}<1\arcsec$ ($\lesssim8000$ au \rr{assuming a distance 8 kpc})
    \item Its chemistry is unlike that of other known objects: the ratio of SO to CS, and SO and CS to other molecules, is different from other Galactic Center and Galactic disk clouds and from evolved stars and stellar merger remnants.  Most significant is the nondetection of SiO, which is a strong indication that there are no shocks in the MUBLO.
    \item It is dusty, as evidenced by the spectral index \rr{of its continuum emission} from \threemm to 850 \um, $\alpha=3.25$.
    \item It is likely massive, with $M_\mathrm{gas}\sim 50 \msun$ assuming typical ISM dust \rr{and a gas-to-dust ratio of 100}, and therefore dense.
    \item It is cold, $T_\mathrm{gas}\sim15$ K and $T_\mathrm{dust} < 50$ K.
    \item \rr{No counterpart is found at other wavelengths for which comparable resolution data exist, including infrared (1-25 \um), radio ($\lambda > 3$ mm), and X-ray (2-10 keV).}
\end{itemize}

Given these observational features, we considered many physical explanations of the MUBLO.
Among the most promising for followup are the stellar merger and intermediate mass black hole hypotheses.
There are no exact analogs to the MUBLO among known astronomical objects.
Future mid-infrared and millimeter observations will be needed to determine what this object is.

\vspace{1cm}

\textbf{Acknowledgements:}
We thank the anonymous referee for a helpful and constructive review.
We thank Brett McGuire, Andres Izquierdo, Joel Kastner, and Elias Aydi for conversations about chemistry, disks, evolved stars, and novae that motivated some of the discussion sections.
AG acknowledges support from the NSF under grants AAG 2008101, 2206511, and CAREER 2142300.
The authors acknowledge University of Florida Research Computing for providing computational resources and support that have contributed to the research results reported in this publication. 
KMD and SV are funded by the European Research Council (ERC) Advanced Grant MOPPEX 833460.vii.
XL acknowledges support from the National Key R\&D Program of China (No.\ 2022YFA1603101), the Natural Science Foundation of Shanghai (No.\ 23ZR1482100), the National Natural Science Foundation of China (NSFC) through grant Nos. 12273090 \& 12322305, and the Chinese Academy of Sciences (CAS) `Light of West China' Program (No.\ xbzgzdsys-202212). LC, IJS, and VMR acknowledge funding from grants No. PID2019-105552RB-C41 and PID2022-136814NB-I00 by the Spanish Ministry of Science, Innovation and Universities/State Agency of Research MICIU/AEI/10.13039/501100011033 and by ERDF, UE. V.M.R. also acknowledges support from the grant number RYC2020-029387-I funded by MICIU/AEI/10.13039/501100011033 and by "ESF, Investing in your future", and from the Consejo Superior de Investigaciones Cient{\'i}ficas (CSIC) and the Centro de Astrobiolog{\'i}a (CAB) through the project 20225AT015 (Proyectos intramurales especiales del CSIC).
RSK acknowledges financial support from the European Research Council via the ERC Synergy Grant ``ECOGAL'' (project ID 855130),  from the German Excellence Strategy via the Heidelberg Cluster of Excellence (EXC 2181 - 390900948) ``STRUCTURES'', and from the German Ministry for Economic Affairs and Climate Action in project ``MAINN'' (funding ID 50OO2206). 
A.S.-M.\ acknowledges support from the RyC2021-032892-I grant funded by MCIN/AEI/10.13039/501100011033 and by the European Union `Next GenerationEU’/PRTR, as well as the program Unidad de Excelencia María de Maeztu CEX2020-001058-M, and support from the PID2020-117710GB-I00 (MCI-AEI-FEDER, UE).
C.\ Battersby  gratefully  acknowledges  funding  from  National  Science  Foundation  under  Award  Nos. 1816715, 2108938, 2206510, and CAREER 2145689, as well as from the National Aeronautics and Space Administration through the Astrophysics Data Analysis Program under Award No. 21-ADAP21-0179 and through the SOFIA archival research program under Award No.  09$\_$0540.  
Q. Zhang acknowledges the support of National Science Foundation under award No. 2206509.

This paper makes use of the following ALMA data: ADS/JAO.ALMA\#2021.1.00172.L, ADS/JAO.ALMA\#2012.1.00080.S, ADS/JAO.ALMA\#2017.1.01185.S. ALMA is a partnership of ESO (representing its member states), NSF (USA) and NINS (Japan), together with NRC (Canada), NSTC and ASIAA (Taiwan), and KASI (Republic of Korea), in cooperation with the Republic of Chile. The Joint ALMA Observatory is operated by ESO, AUI/NRAO and NAOJ.
The National Radio Astronomy Observatory is a facility of the National Science Foundation operated under cooperative agreement by Associated Universities, Inc.

\clearpage
\appendix
\section{Total Power Spectra}
\label{appendix:totalpowerspectra}
\begin{figure}
    \centering
    \includegraphics[width=0.49\textwidth]{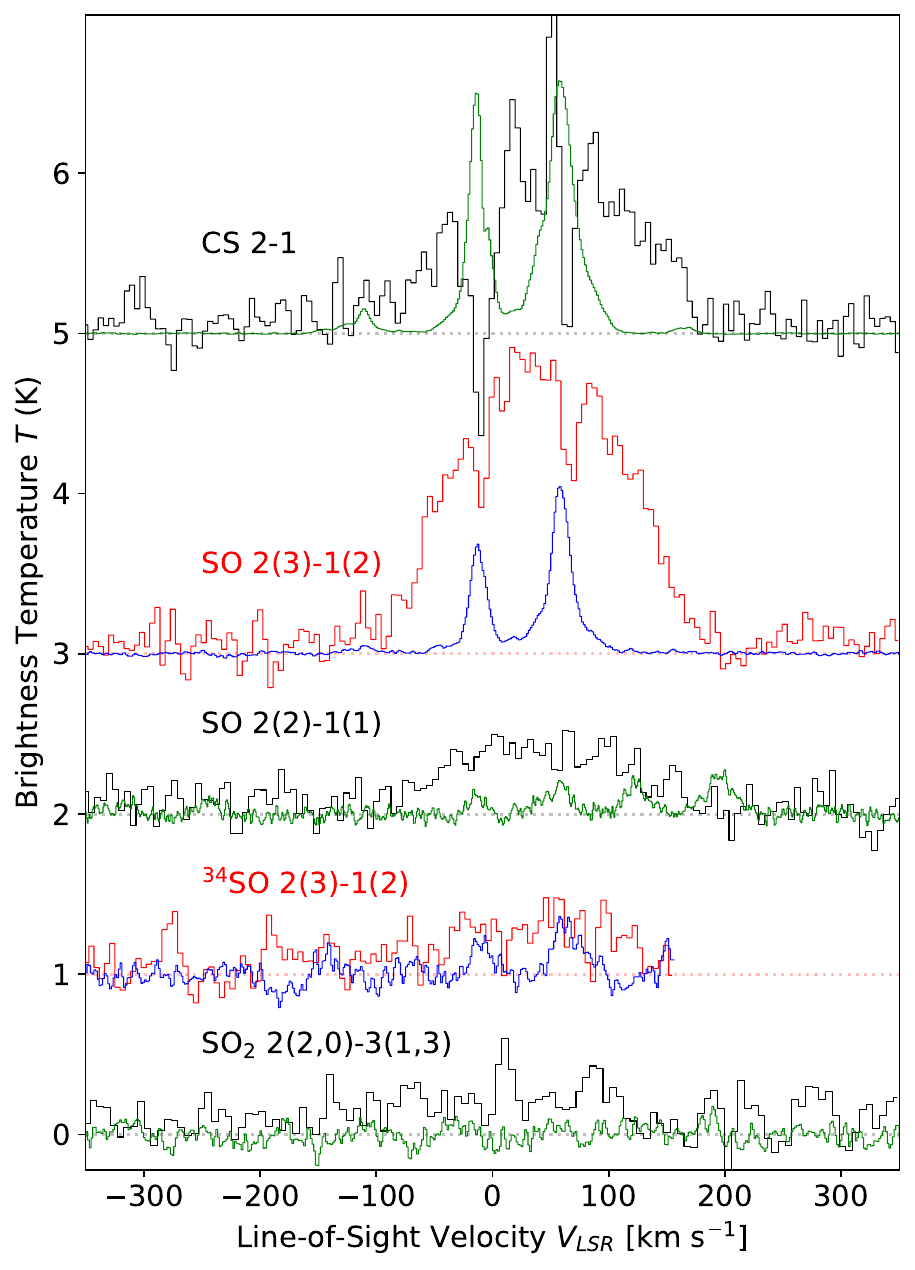}
    \includegraphics[width=0.49\textwidth]{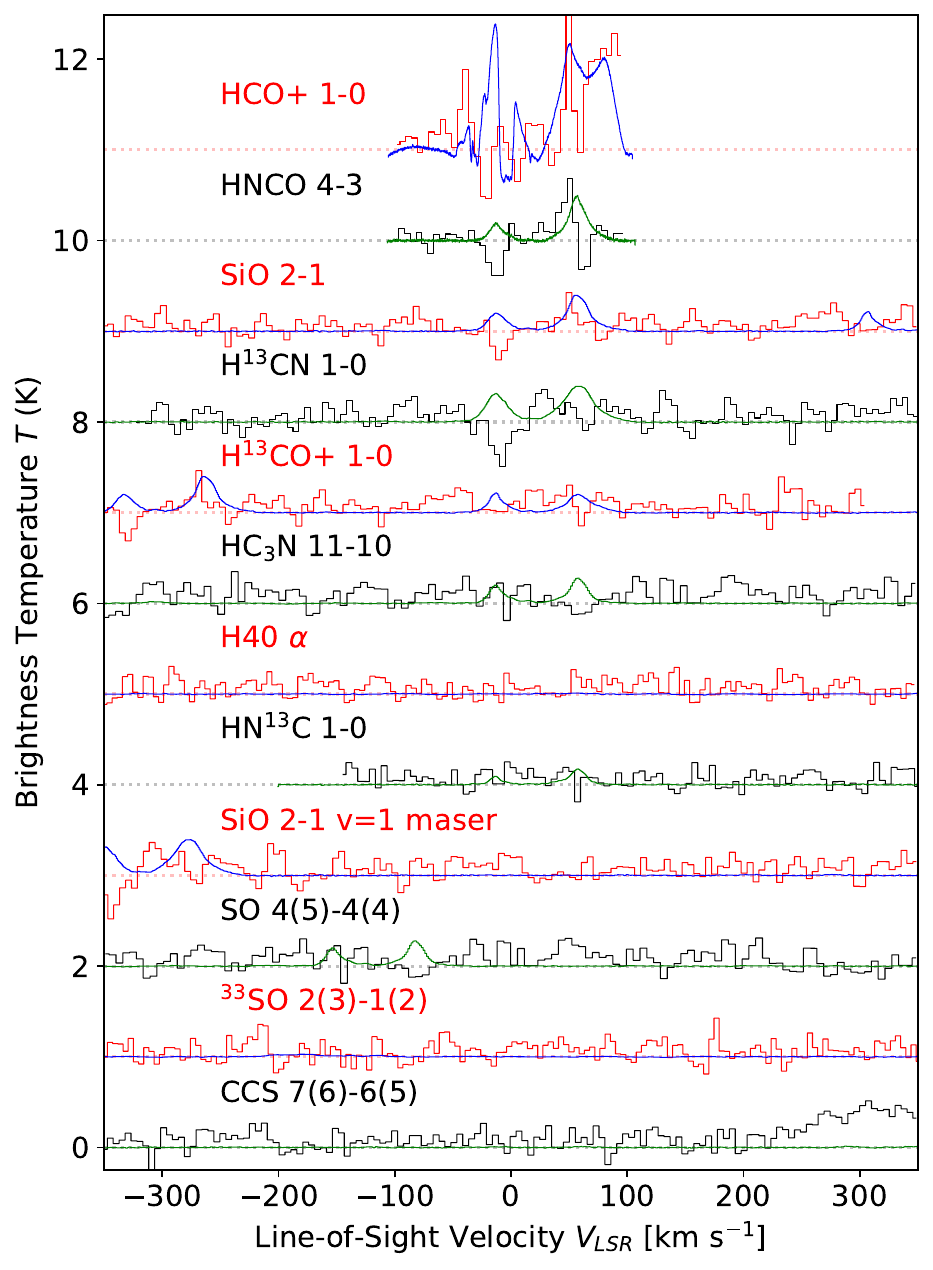}
    \caption{A repeat of Figure \ref{fig:coarse_spectra} with the Total Power (TP) data at this pointing, with $\sim1'$ resolution, overlaid.
    \rr{For the detections (left), the TP data are scaled by factors of $1, 3, 10, 20, 20$ for CS, SO 2(3), SO 2(2), $^{34}SO$ 2(3), and SO$_2$, respectively.}
    The overlap between the TP peaks and the 12m array data dips confirms that the narrow features come from foreground molecular clouds.
    }
    \label{fig:coarse_spectra_withTP}
\end{figure}

Figure \ref{fig:coarse_spectra_withTP} shows the ACES 12m spectra with spectra from the total power (TP; 12m single-dish) array overlaid.
The TP peaks coincide with the dips in the 12m spectra, indicating that the dips are caused either by absorption lines from foreground clouds or from interferometric imaging artifacts from resolved-out clouds.
It is not possible to tell from the images whether the absorption is real or not, despite the clear absorption-line profile seen in Figure \ref{fig:fittedspectra}.

\section{Chemical Modeling}
\label{appendix:chemistry}

To guide our interpretation of the chemistry of the MUBLO, we use two types of astrochemical modeling codes: Meudon PDR \citep{LePetit2006} and UCLCHEM \citep{Holdship2017}. The former is a photochemical model of a one-dimensional stationary PDR and the latter is a zero-dimensional model that tracks the time evolution of \rr{the chemistry of} a cloud.

The Meudon PDR code computes abundances of atoms and molecules and level excitation of any number of species at each position in the cloud (a stationary plane-parallel slab of gas and dust illuminated by radiation field). 
The code solves the FUV radiative transfer at each point of the cloud, taking into account absorption in the continuum by dust and in discrete transitions by H and H$_2$.
This model computes the thermal balance taking into account heating processes (cosmic rays, photoelectric effect on grains, H$_2$ formation on grains, chemistry, grain-gas coupling, turbulence,  photons: photodissociation and photoionization, and secondary photons) and cooling from discrete radiative transitions in IR and mm lines of various species following collisional excitation, free-free emission and H$_2$ dissociation.
A post-processor code computes the line intensities and column densities. 

The UCLCHEM code is a zero-dimensional model that follows the time evolution of \rr{the chemistry of} a cloud.
We use it to predict chemical abundances rather than intensities.

\subsection{FUV and CR irradiation models}
\label{sec:meudon}
We use the Meudon PDR code to model the MUBLO as a constant density cloud illuminated by the FUV radiation field ($G_0$).
We adopt enhanced (compared to that in Galactic disk GMCs) H$_2$ cosmic-ray ionization rates \citep[e.g., $\zeta_\mathrm{CR}$$\approx$(3.5$\pm$1.4)$\times$10$^{-16}$s$^{-1}$ for the LSR velocity range 20-75\,\kms associated with the 50 km s$^{-1}$ cloud, see][]{Indriolo2015}.
Table \ref{tab:MeudonPDR} shows the input parameters. We accounted for sulfur depletion usually needed in starless cores (a factor $>$100) and that estimated in hot cores or bipolar outflows \citep[a factor $\sim$10; see][and references therein]{Fuente2023}.

\begin{figure}[ht]
    \centering
    \includegraphics[width=0.32\textwidth]{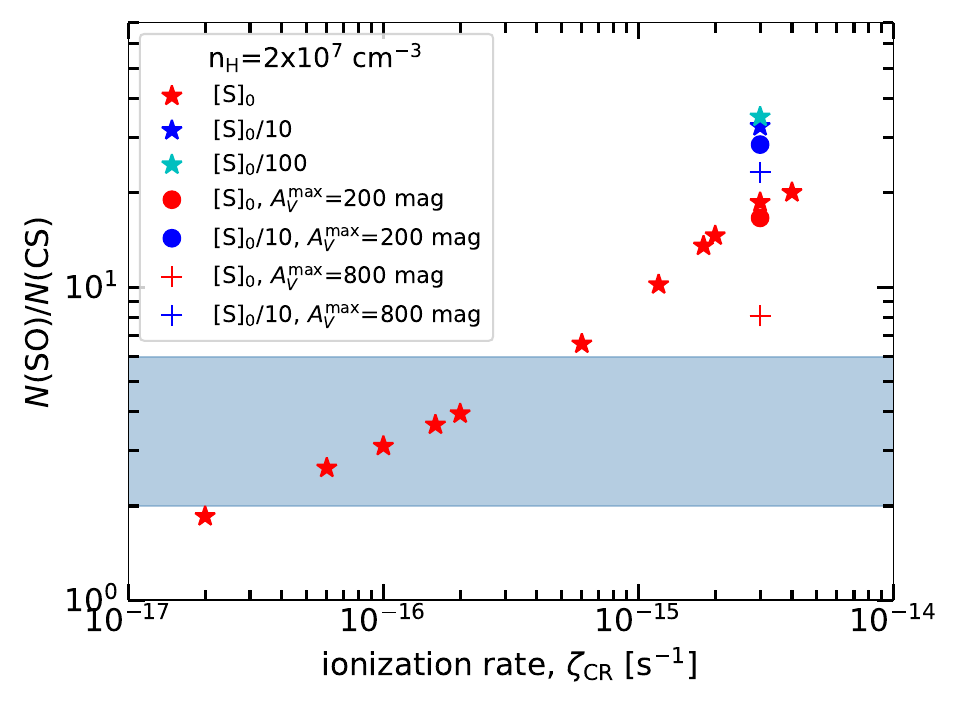}\hspace{0.02in}
    \includegraphics[width=0.32\textwidth]{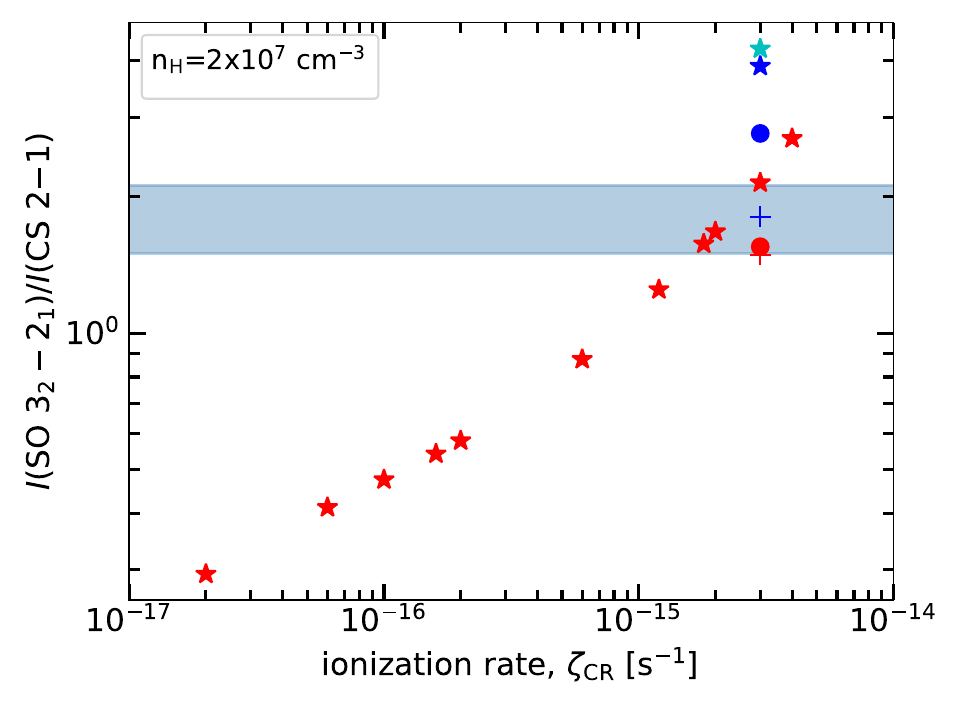} \hspace{0.02in}
    \includegraphics[width=0.32\textwidth]{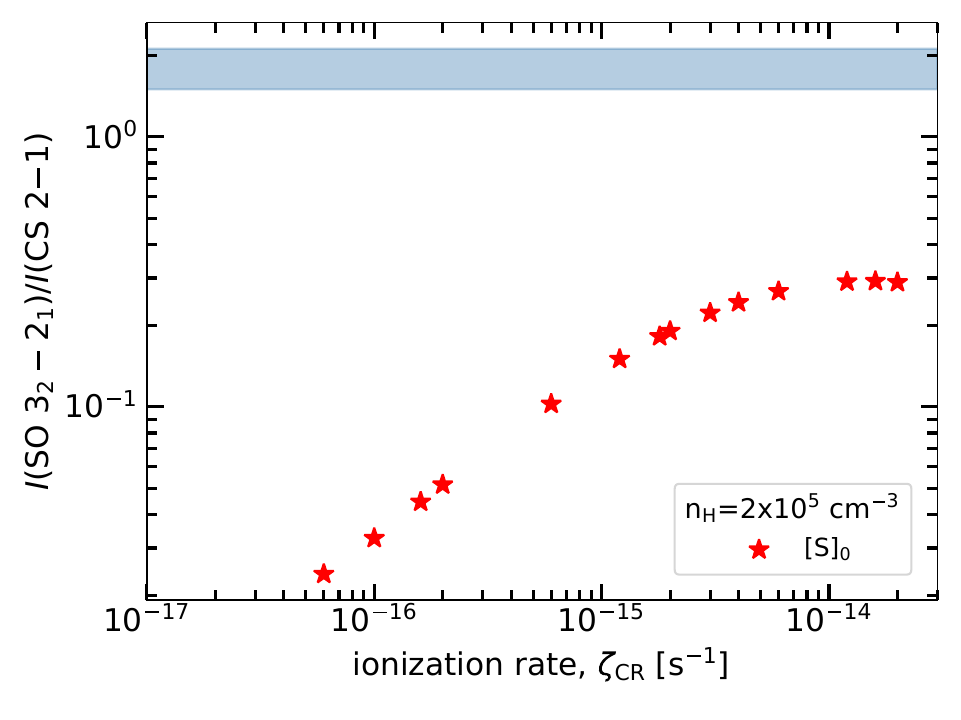}
    \caption{Meudon PDR model predictions for different values of $\zeta_\mathrm{CR}$ (and fixed $G_0$=10$^3$, $n_\mathrm{H}$=2$\times$10$^7$~cm$^{-3}$ (left and center),  $n_\mathrm{H}$=2$\times$10$^5$~cm$^{-3}$ (right) 
    , and $A_\mathrm{V, max}$=20, 200, and 800~mag).  Blue and cyan stars are the ratios in models with sulfur depleted 10 and 100 times. Red and blue dots are models with $A_\mathrm{V}^\mathrm{max}$=200~mag and sulfur abundance [S]$_0$ and [S]$_0$/10, respectively. Red and blue crosses are models with $A_\mathrm{V}^\mathrm{max}$=800~mag and sulfur abundance [S]$_0$ and [S]$_0$/10, respectively.   \textit{Left}: Markers show the SO vs CS column density ratio. The blue horizontal shaded area represents the estimated column density ratio $\sim$4 with a 50\% uncertainty (see Sect.~\ref{sec:LTE}). \textit{Center} \rr{and \textit{right}}: Markers show the SO~3$-$2 to CS~2$-$1 intensity ratio. The blue horizontal shaded area represents the observed intensity ratio $\sim$1.8 (see Table \ref{tab:spectral_measurements}) with a 20\% uncertainty.} 
    \label{fig:SO-CS_Meudon}
\end{figure}

Figure~\ref{fig:SO-CS_Meudon} shows the model predictions of the SO to CS column density ratio, \mbox{$N$(SO)/$N$(CS)}, and the \mbox{$I$(SO 3$_2$--2$_1$)/$I$(CS 2$-$1)} intensity ratio as function of the cosmic ray ionization rate ($\zeta_\mathrm{CR}$).
The blue shaded area represents the estimated and observed ratios, respectively (see Sect.~\ref{sec:LTE} and Table~\ref{tab:spectral_measurements}).
\rr{The estimated region comes from LTE modeling (\S \ref{sec:LTE}), while the Meudon model performs non-LTE modeling of the line ratio, which is why the blue shaded regions are different for these two panels - i.e., the Meudon model suggests that the line ratio is not in LTE.}
These models show that a high density ($n_\mathrm{H}\gtrsim10^7$ \percc) is required to match the observations: when $n_\mathrm{H}$=2$\times$10$^5$~\percc, \mbox{$N$(SO)/$N$(CS)} is larger than 20 for all $\zeta_\mathrm{CR}$.
Hence, the computed \rr{intensity and column density ratios} never overlap with the estimated ratio \rr{from} Sect.~\ref{sec:LTE}.

\begin{figure*}[!ht]
    \centering
    \includegraphics[width=0.45\textwidth]{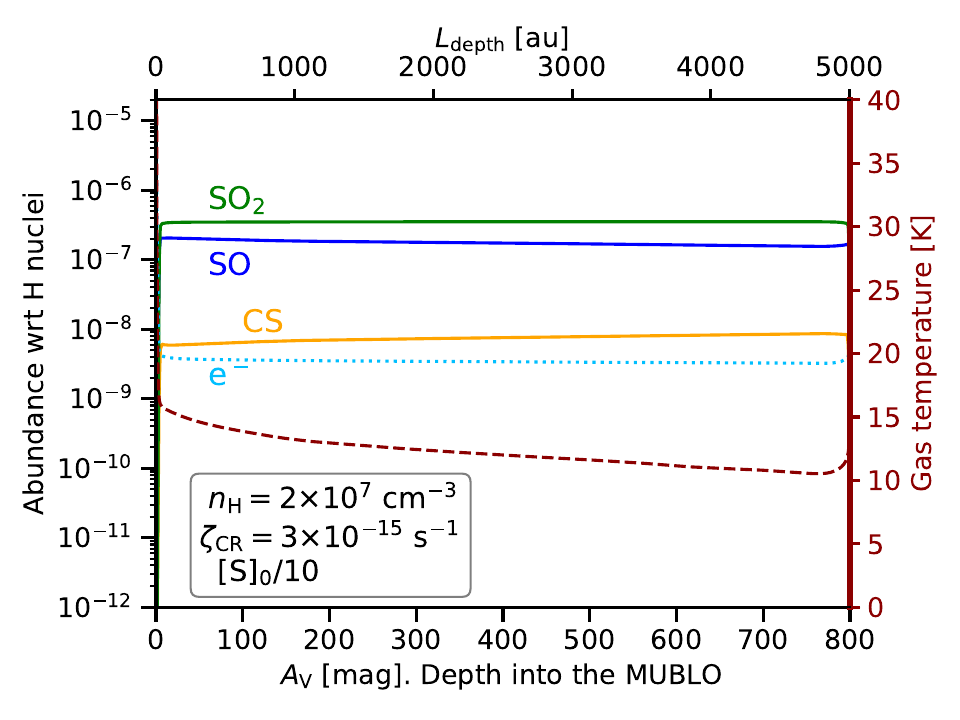}
    \caption{Abundance and $T_\mathrm{k}$ profiles of constant density gas-phase photochemical models with $n_\mathrm{H}$=2$\times$10$^{7}$~\percc, $\zeta_\mathrm{CR}$=3$\times$10$^{-15}$~s$^{-1}$, and sulfur abundance depleted a factor 10 (see blue cross in Fig.~\ref{fig:SO-CS_Meudon}).}
    \label{fig:AV_SO-CS_Meudon}
\end{figure*}

Figure~\ref{fig:AV_SO-CS_Meudon} shows the abundance and gas temperature profiles of a model with $n_\mathrm{H}$=2$\times$10$^{7}$~\percc, $\zeta_\mathrm{CR}$=3$\times$10$^{-15}$~s$^{-1}$ per H$_2$ molecule, and sulfur abundance depleted a factor 10.
The gas temperature at large $A_\mathrm{V}$ is $T_\mathrm{k}$$\sim$10-15~K, which agrees with the measured $T_{LTE}=T_{ex}(\mathrm{SO})$.

In Sect.~\ref{sec:spectralmeasurements} we show that this source has $I$(SO 3$_2$--2$_1$)/$I$(CS 2$-$1) ratio of about 1.8, while it has $N$(SO)/$N$(CS) of about 4, though the estimated $N$(CS) value is dominated by systematic uncertainty in $T_\mathrm{ex}$. 
The model predictions result in a narrow range of parameters consistent with the observed SO/CS intensity ratio and estimated SO/CS column density ratio, $n_\mathrm{H}$$\sim$2$\times$10$^7$~\percc and 10$^{-16}$$\lesssim$$\zeta_{CR}$$\lesssim$5$\times$10$^{-15}$~s$^{-1}$.

\begin{table*}[ht]
    \begin{center}
    \caption{Input parameters in the Meudon PDR code}
    \begin{tabular}{llcc}
       \toprule
       Parameter  & Unit & Value & Note  \\
       \midrule
       $n_\mathrm{H}$  & cm$^{-3}$ & 2$\times$10$^5$,
       2$\times$10$^7$  & $n_\mathrm{H}$=$n$(H)+2$n$(H$_2$), (1)  \\ 
       $A_\mathrm{V,max}$ & mag & 20, 200, 800 & (2), $L$$\sim$125, 1250, 5000 au \\
       FUV radiation field, $G_0$ & -- & 10$^3$ & (3) \\
       Cosmic ray ionization rate, $\zeta_\mathrm{CR}$ & s$^{-1}$ per H$_2$ &[2$\times$10$^{-17}$, 2$\times$10$^{-14}$] & (4), (5)   \\
       Dust extinction curve & & Galaxy & (6) \\ 
       $R_\mathrm{V}$ & & 3.1 & (6) \\
       $N_\mathrm{H}$/$E$(B-V) = $C_\mathrm{D}$ & cm$^{-2}$ & 5.8$\times$10$^{21}$ & (7) \\ 
       $\left [\mathrm{S}  \right ]$ & -- & $\left [\mathrm{S}  \right ]_0  \times$ (1, 0.1, 0.01)  & $\left [\mathrm{S}  \right ]_0$=1.4$\times$10$^{-5}$ (8), (9), (10) \\
       
       \bottomrule       
        \label{tab:MeudonPDR}
    \end{tabular}
    \end{center}
    \par \vspace{0.05in}
    (1) See Sect.\ref{sec:nonlte}; (2) The relation between $A_\mathrm{V, max}$ and the size of the computed cloud, $L$, in cm is \mbox{$L=N_\mathrm{H}/n\mathrm{_H}=(C_\mathrm{D}/n_\mathrm{H})\,(A_\mathrm{V}/R_\mathrm{V})$}; (3) Scaling parameter to the ISRF, \citet{Mathis1983}; (4) \citet{Indriolo2015}; (5) \citet{LePetit2016}; (6) \citet{Fitzpatrick2007}; (7) \citet{Bohlin1978};  (8) \citet{Asplund2009}; (9) \citet{Goicoechea2021}; (10) \citet{Fuente2023}.
    \par
\end{table*}


\clearpage
\subsection{UCLCHEM models}
\label{sec:uclchem}
\begin{figure}[!ht]
    \centering
    \includegraphics[width=1.0\textwidth]{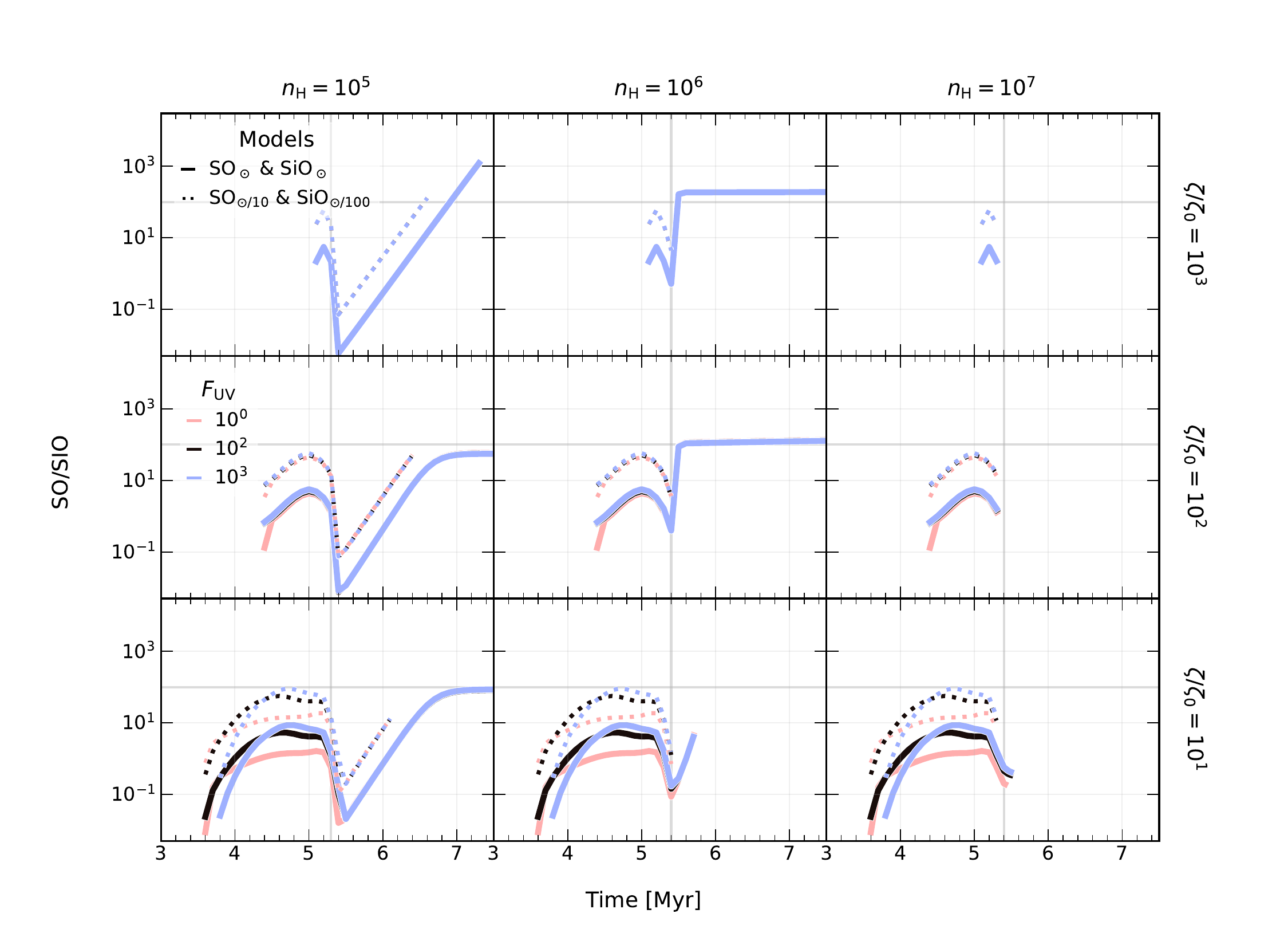}
    \caption{The evolution of SO/SiO ratio in UCLCHEM models at a temperature of 15 K for both gas and dust. The parameters used in the models are detailed in Table \ref{tab:UCLCHEM}. The ratio value of 100, which is the lower limit derived in \S \ref{sec:LTE}, is represented by a thick horizontal gray line, while the vertical line indicates the time when the final density is reached. The ratio was calculated for time steps where both species are above the observable limit, i.e., their abundance is $\geq 10^{-12}$. For all models where the ratio of SO / SiO $> 100$, that ratio only occurs at late times, after the final density is reached. However, no models with the density of $10^7\,\mathrm{cm}^{-3}$, in the rightmost column, can predict the observed ratio.}
    \label{fig:UCLCHEM-15}
\end{figure}

\begin{figure}[!ht]
    \centering
    \includegraphics[width=1.0\textwidth]{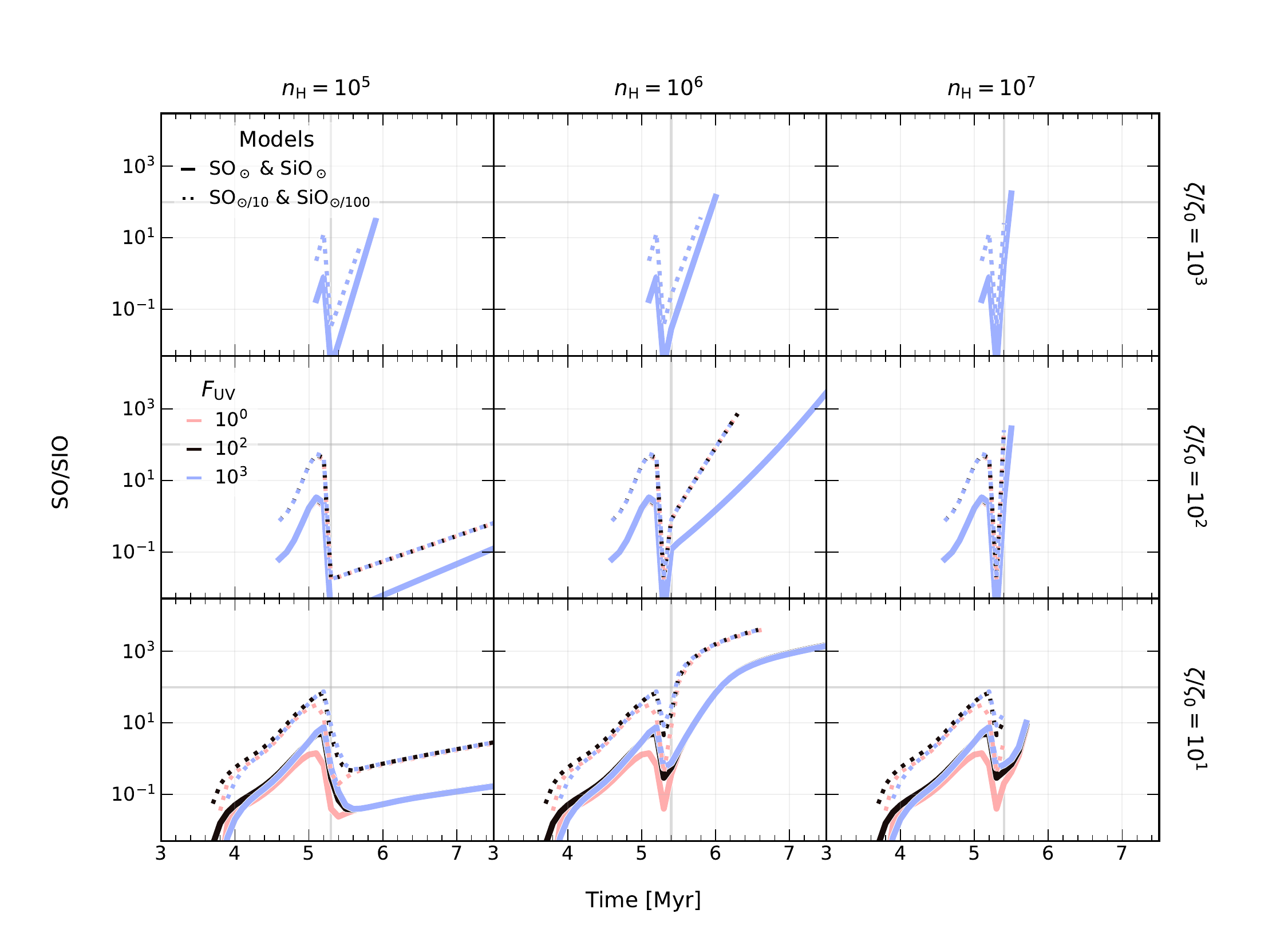}
    \caption{As in Fig. \ref{fig:UCLCHEM-15}, but for a temperature of 50 K. In this case, the ratio of SO/SiO is greater than 100 for all densities considered. However, for a density of $10^5\,\mathrm{cm}^{-3}$, it takes a much longer \rr{time} to reach the desired ratio, which exceeds the presented time range. }
    \label{fig:UCLCHEM-50}
\end{figure}

We describe the UCLCHEM models run in this section.
Table \ref{tab:UCLCHEM} shows the parameter grid we ran.
We accounted for the observed depletion of silicon \citep[Si; e.g.,][]{Savage1996} and sulfur \citep[S; e.g.,][]{Palumbo1997} abundances by evaluating both a model with solar initial abundances and a model with a depletion of a factor 10 for S and 100 for Si.
For models with gas and dust temperatures set to 15 K (see Fig. \ref{fig:UCLCHEM-15}), we find that the observed ratio SO/SiO~$\gtrsim100$ is predicted for  densities $\leq10^6\,\mathrm{cm}^{-3}$, while for the 50 K models (see Fig. \ref{fig:UCLCHEM-50}), the ratio can be recovered for all considered densities.
SO/SiO$>100$ is reachable for all considered cosmic ray ionization rates and UV field strengths regardless of the gas and dust temperatures, however, in the case of the lowest density scenario ($n=10^5$ \percc) for the 50 K models, the ratio only is recovered after more than $10^7$ years.
For the medium density scenario, the ratio is quickly achieved and persists over a long chemical timescale. For the highest density scenario, the ratio increases rapidly during the collapse, but once the final density is reached both SiO and SO quickly freeze out.

Moreover, \cite{DutkowskaVermarien2024} explored a wide range of continuous shock model parameters, covering shock velocity $5-30\,\mathrm{km\,s}^{-1}$, pre-shock medium temperature $15-35$ K, pre-shock density $10^4-10^6\,\mathrm{cm}^{-3}$, magnetic field $10^{1}-10^{3}\,\mu\mathrm{G}$, cosmic ray ionization rate $1.31\times10^{-16}-1.31\times10^{-13}\,\mathrm{s}^{-1}$, and UV irradiation $10^0-10^4$ Habing. 
In these models, SO/SiO ratios are low, $<100$, during and shortly after the shock.
The high SO/SiO ratio in the MUBLO is therefore either inherited from the collapse/quiescent cloud stage, or it is achieved long after the shock has passed. 
Shocks with a velocity of $\leq 5\,\mathrm{km\,s}^{-1}$ are the only ones capable of maintaining a ratio above 100. 
For a higher shock velocity, the SO/SiO ratio stays below 100, though given enough time after the shock, the SO/SiO ratio can eventually exceed 100 again.

Lastly, models at higher temperatures of $100$K and $150$K were simulated, since these higher temperatures are observed in the CMZ \citep{ginsburgDenseGasGalactic2016,millsDetectionWidespreadHot2013,zengComplexOrganicMolecules2018a}.
For these models an identical grid of models was run. These models reproduce the SO/SiO ratio only for high cosmic ray ionization rates, under which conditions the SiO is quickly destroyed and the SO reaches a steady state.

From these time dependent gas-grain simulations, we conclude that a shock-free medium is favored to reproduce the high SO/SiO ratios.
The ratios can be reproduced sufficiently by dense and possibly quiescent environments.

\begin{table*}[ht]
\caption{Parameter space covered with UCLCHEM models}
\label{tab:UCLCHEM}
\centering
\begin{tabular}{l|l|l|l}
\hline
\hline
Parameter                         & Unit      & Values             & Notes                                                                                              \\
\hline
$n_\mathrm{H,\,final}$            & cm$^{-3}$ & $10^5, 10^6, 10^7$ & $n_\mathrm{H} = n(\mathrm{H}) + 2n(\mathrm{H}_2)$; $n_\mathrm{H,\, init} = 10^2\,\mathrm{cm}^{-3}$ \\
$T_\mathrm{gas,\,dust}$           & K         & $15, 50$           & ...                                                                                                \\
$\zeta / \zeta_0$                 & $-$       & $10^1, 10^2, 10^3$    & $\zeta_0 = 1.310\times10^{-17}\,\mathrm{s}^{-1}$                                                           \\
$F_\mathrm{UV}$                   & Habing    & $10^0, 10^2, 10^3$     & ...                                                                                                \\
$X(\mathrm{Si})$    & $-$      &$1.78^{-8}, 1.78^{{-6}\dag}$ & $^\dag X(\mathrm{Si})_\odot$ \citep{Jenkins2009}  \\
$X(\mathrm{S})$     & $-$     & $3.51^{-7}, 3.51^{{-6}\dag}$     & $^\dag X(\mathrm{S})_\odot$ \citep{Jenkins2009}         \\
\hline
\end{tabular}
\end{table*}

\section{RADEX models}
\label{appendix:radex}
\begin{figure}
    \centering
    \includegraphics[width=0.45\textwidth]{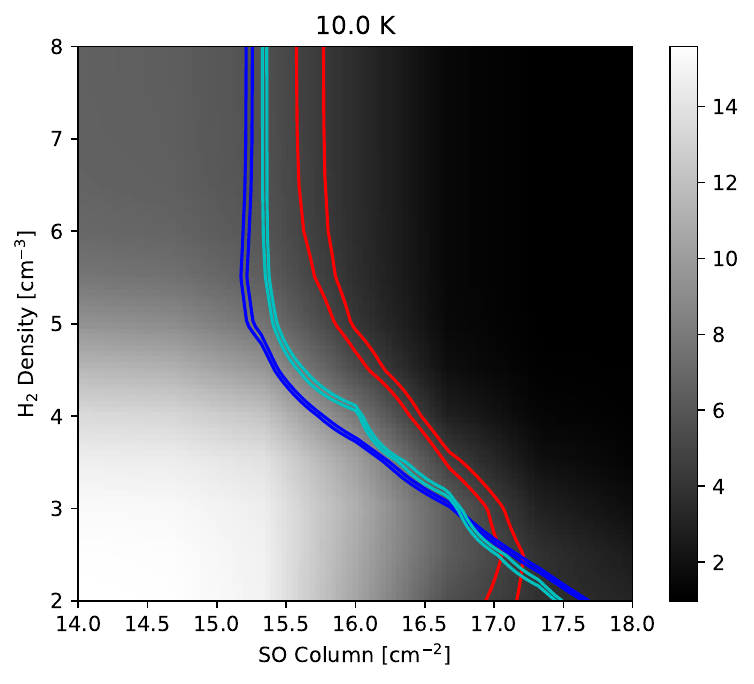}
    \includegraphics[width=0.45\textwidth]{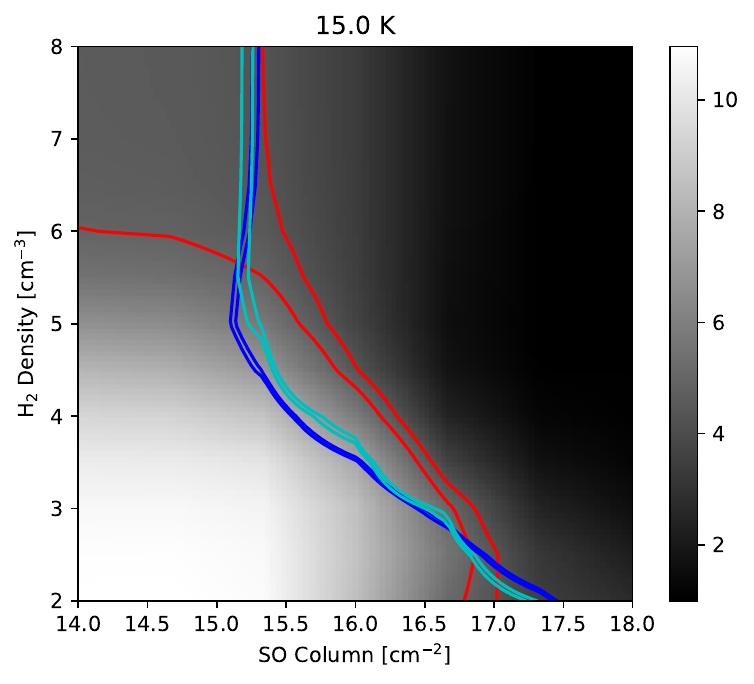}
    \includegraphics[width=0.45\textwidth]{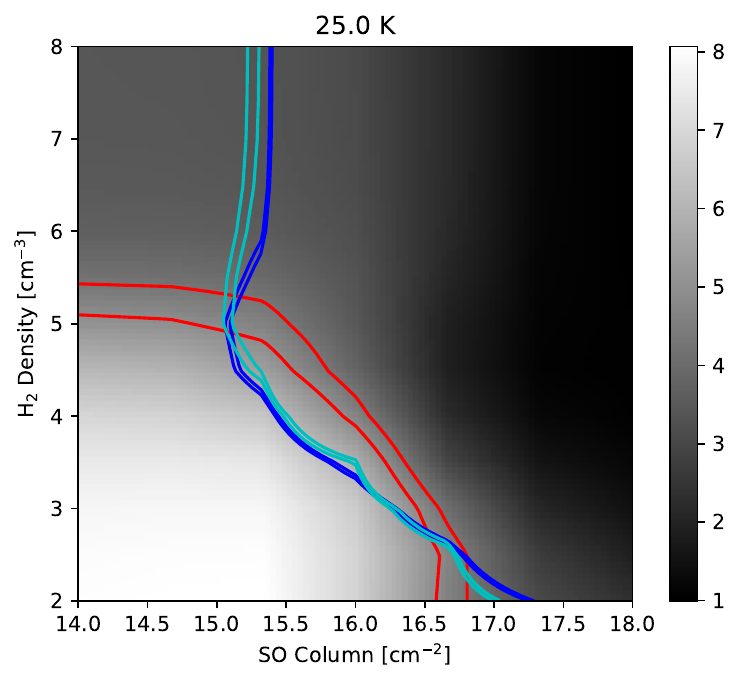}
    \includegraphics[width=0.45\textwidth]{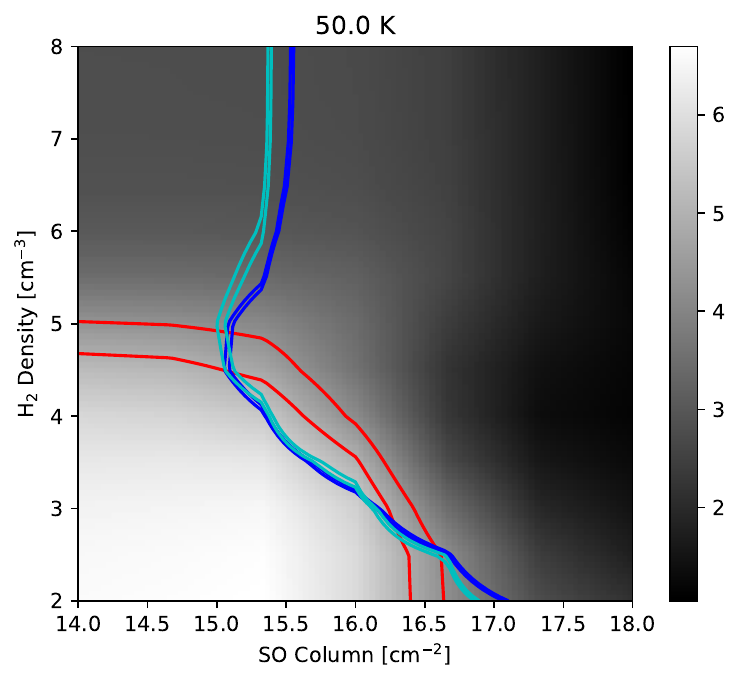}
    \caption{RADEX model grids used to constrain the temperature and density.
    Each plot shows in grayscale the ratio of the surface brightness of SO 2(3)-1(2) to SO 2(2)-1(1).
    The red contour shows the observationally allowed range, $R_{SO} = 4.87 \pm 0.33$.
    The cyan and blue contours show the allowed values for SO 2(2)-1(1) and SO 2(3)-1(2), respectively: $S_{2-1}=0.38\pm0.02$ K and $S_{3-2}=1.87\pm0.03$ K.
    The parameter space up and to the right (higher density and column density) of the blue curves and within the red curves is also allowed if the filling factor is $ff<1$.
    }
    \label{fig:radexgrids}
\end{figure}

We show parameter slices from the grid of RADEX models described in \S \ref{sec:nonlte}.
The models use the large velocity gradient (LVG) approximation with $dv=70$ \kms.
Figure \ref{fig:radexgrids} shows these model grids.

\bibliographystyle{aasjournal}

\end{document}